\documentclass[twocolumn,notitlepage,nofootinbib,floatfix,superscriptaddress]{revtex4-2}
\usepackage{amsmath, amsfonts, amsthm, amssymb, bbm}
\usepackage[margin=1in]{geometry}
\usepackage{algorithm,algpseudocode}
\usepackage{graphicx,xcolor,tikz}
\usepackage{hyperref}
\usepackage{physics}
\usepackage{qcircuit}

\newcommand{\mc}{\mathcal}

\newcommand{\unit}[1]{\,\mathrm{#1}}

\algblockdefx[ForEach]{ForEach}{EndForEach}[1]{\textbf{for each} #1 \textbf{do}}{\textbf{end for each}}

\newcommand{\measureX}[1]{*+[F-:<.9em>]{#1}}

\begin{document}

\title{Optimizing quantum error correction protocols with erasure qubits}
\author{Shouzhen Gu} 
\thanks{These authors contributed equally.}
\affiliation{California Institute of Technology, Pasadena, CA 91125, USA}
\affiliation{
	Yale Quantum Institute \& Department of Applied Physics,
	Yale University, New Haven, CT 06520, USA}
\author{Yotam Vaknin}
\thanks{These authors contributed equally.}
\affiliation{AWS Center for Quantum Computing, Pasadena, CA 91125, USA}
\affiliation{Racah Institute of Physics, The Hebrew University of Jerusalem, Jerusalem, 91904, Givat Ram, Israel}
\author{Alex Retzker}
\affiliation{AWS Center for Quantum Computing, Pasadena, CA 91125, USA}
\affiliation{Racah Institute of Physics, The Hebrew University of Jerusalem, Jerusalem, 91904, Givat Ram, Israel}
\author{Aleksander Kubica}
\affiliation{AWS Center for Quantum Computing, Pasadena, CA 91125, USA}
\affiliation{California Institute of Technology, Pasadena, CA 91125, USA}
\affiliation{
	Yale Quantum Institute \& Department of Applied Physics,
	Yale University, New Haven, CT 06520, USA}
\date{\today}

\begin{abstract}
	Erasure qubits offer a promising avenue toward reducing the overhead of quantum error correction (QEC) protocols.
	However, they require additional operations, such as erasure checks, that may add extra noise and increase runtime of QEC protocols.
	To assess the benefits provided by erasure qubits, we focus on the performance of the surface code as a quantum memory.
	In particular, we analyze various erasure check schedules, find the correctable regions in the phase space of error parameters and probe the subthreshold scaling of the logical error rate.
	We then consider a realization of erasure qubits in the superconducting hardware architectures via dual-rail qubits.
	We use the standard transmon-based implementation of the surface code as the performance benchmark.
	Our results indicate that QEC protocols with erasure qubits can outperform the ones with state-of-the-art transmons, even in the absence of precise information about the locations of erasure errors.
\end{abstract}

\maketitle

\section{Introduction}
Quantum error correction (QEC) with erasure qubits has been a promising avenue toward reducing the overhead associated with fault-tolerant quantum computation~\cite{wu2022erasure,Kang2023,kubica2023heralding,Kubica2023,Teoh2023,gu2023faulttolerant,ma2023high,Scholl2023,Chou2023,Levine2023,koottandavida2024erasure}.
In an erasure qubit, the dominant noise removes the state from the computational subspace. Information from the detection of such errors, which we refer to as erasures, can be effectively used to improve the performance of QEC protocols.
While the concept of erasures via the dual-rail construction had a central role in photonic quantum computation~\cite{chuang1995simple,knill2001scheme,knill2003bounds,bartolucci2023fusion,silva2005thresholds}, various realizations of QEC with erasure qubits have recently been proposed for other platforms, including neutral atoms~\cite{wu2022erasure}, trapped ions~\cite{Kang2023} and superconducting circuits~\cite{kubica2023heralding,Kubica2023,Teoh2023}, with several promising experimental demonstrations~\cite{ma2023high,Scholl2023,Chou2023,Levine2023,koottandavida2024erasure}.  
Erasure qubits can offer improvements for arbitrary QEC codes, although recent works have focused on quantifying the benefits for topological codes, such as the surface code~\cite{Dennis2002} or the honeycomb code~\cite{Hastings2021}.

Compared to standard QEC protocols, using erasure qubits comes at a cost.
Namely, quantum circuits need to include additional elements, erasure checks (ECs) and reset operations, that are capable of diagnosing erasures and reinitializing erased qubits back to the computational subspace.
These operations take additional time during which noise may accumulate.
From this perspective, one would prefer to implement as few of them as possible.
On the other hand, ECs provide valuable information about the location of possible erasures.
Therefore, one expects an optimal frequency of ECs that balances these two effects.
Most of previous works have assumed that ECs are performed after every entangling operation without optimizing their frequency.

In this article, we assess the benefits for QEC protocols provided by erasure qubits.
Using the framework for QEC with erasure qubits introduced in Ref.~\cite{gu2023faulttolerant}, we analyze the performance of the surface code as a quantum memory.
For various EC schedules, we obtain the correctable region in the phase space of error parameters; see Fig.~\ref{fig:thresholdsurfaces}.
We also analyze the subthreshold scaling of the logical error rate.
In our simulations, we use a novel decoding method that relies on an approximate conversion of QEC protocols with erasure qubits into stabilizer circuits, which may be of independent interest. 

We then benchmark the performance of QEC protocols with transmon dual-rail qubits, which are erasure qubits composed of two coupled transmons~\cite{campbell2020universal, Kubica2023, Levine2023}, against the standard $T_1$-limited tramsmons.
We observe that using dual-rail qubits results in an approximately 50\% increase in tolerable physical error rates (i.e., a 50\% increase in the QEC threshold),
provided that the measurement times of dual-rail qubits and transmons are comparable and there is a significant bias between the amplitude damping noise and Pauli errors.
A key factor that indicates the advantageous performance of dual-rail qubits is the ratio of the measurement time $T_M$ to the two-qubit (2Q) gate time $T_{2Q}$---the smaller the ratio $T_M/T_{2Q}$, the better for dual-rail qubits.
We also analyze different implementations of reset operations.
We compare a \emph{one-way} pulse, which brings the erased state back to the computational subspace without dephasing the computational subspace, with a \emph{unitary} pulse, which transforms between the erased state and the computational subspace.
We find that the most effective protocol employs a one-way pulse exclusively to the qubits identified as erased, and it reduces the impact of measurement errors by approximately a factor of 2.
Lastly, we consider how the gate implementation on dual-rail qubits can be designed to introduce noise bias, and demonstrate how the XZZX surface code can exploit this bias to further increase the error-correcting threshold.

The remainder of the paper is organized as follows.
In Sec.~\ref{sec:prelim}, we review the formalism for QEC protocols with erasure qubits and describe the surface code as a quantum memory.
In Sec.~\ref{sec:generalresults}, we describe an approximate solution to the decoding problem with erasure qubits, as well as analyze the threshold surface and subthreshold performance of the different EC schedules.
While the results of Sec.~\ref{sec:generalresults} are applicable to any implementation of erasure qubits, Sec.~\ref{sec:superconducting} focuses on superconducting qubits.
We compare dual-rail qubits with transmons, and also analyze the different reset operations.
Possible future directions are discussed in Sec.~\ref{sec:discussion}.

\section{Preliminaries}
\label{sec:prelim}
We first review the formalism to describe QEC protocols with erasure qubits.
Then, we explain how to implement the surface code with erasure qubits.
\subsection{Erasure qubits}
\label{subsec:erasurequbitsoverview}

We use the formalism introduced in Ref.~\cite{gu2023faulttolerant} to describe and simulate QEC protocols with erasure qubits. Compared to standard qubits, two additional operations are used when working with erasure qubits: ECs that measure if a qubit is erased, and reset operations that reinitialize a qubit in the computational subspace. For our purposes, a stabilizer circuit is one that consists of the following basic operations: single-qubit (1Q) state preparation in an eigenstate of a Pauli operator, 1Q readout in a Pauli basis, 1Q Clifford gates, and 2Q controlled-Pauli gates. When a stabilizer circuit is implemented with erasure qubits and enhanced with the two additional operations, we call it an \emph{erasure circuit}.

\begin{table}[htpb]
	\begin{tabular}{| p{20mm} | c | c |}
		\hline
		{\bf operation} & {\bf ideal} & {\bf simulated} \\
		\hline\hline
		state \mbox{preparation} & \Qcircuit @C=.8em @R=.7em {
			\lstick{\ket \psi} & \qw} &
		\Qcircuit @C=.8em @R=.7em {
			\lstick{\ket \psi} & \measure{\mc P\!\left(p\right)} & \qw
		} \\
		\hline
		readout & \hspace*{-5mm}\Qcircuit @C=.8em @R=.7em {
			& \meter_{\qquad\qquad\ P}
		} &
		\Qcircuit @C=.8em @R=.7em {
			& \gate{\mathrm{EC}} & \meter_{\qquad\qquad\qquad\ P} &\\
			& \measureX{\mc N(q)} \cwx & \cw & \cw \\
			& & \measureX{\mc N(q)} \cwx[-2] & \cw }\raisebox{-18.3mm}{ } \\
		\hline
		erasure check with reset & \Qcircuit @C=.8em @R=.7em {
			& \gate{\mathrm{EC^*}} & \qw
		} &
		\Qcircuit @C=.8em @R=.7em {
			& \gate{\mathrm{EC}} & \gate{\mathrm R}& \measure{\mathcal P(p)} & \qw \\
			& \measureX{\mathcal N(q)} \cwx & \cw & \cw & \cw}\raisebox{-11mm}{ } \\
		\hline
		1Q gate & \Qcircuit @C=.8em @R=.7em {
			& \gate{G} & \qw}
		& \Qcircuit @C=.8em @R=.7em {
			& \gate{G} & \measure{\mc P(p)} & \qw}\raisebox{-8mm}{ } \\
		\hline
		2Q gate & \Qcircuit @C=.8em @R=.7em {
			& \multigate{1}{G} & \qw\\
			& \ghost{G} & \qw}
		& \Qcircuit @C=.8em @R=.7em {
			& \multigate{1}{G} & \multimeasure{1}{\mc P(p)} & \qw\\
			& \ghost{G} & \ghost{\mc P(p)} & \qw}\raisebox{-8mm}{ } \\
		\hline
	\end{tabular}
	\caption[Adding noise to an ideal circuit for simulation.]{
		Mapping of an ideal circuit to a simulated circuit.
		The error channel $\mc P(p)$ is the 1Q or 2Q depolarizing channel with error rate $p$, and $\mc N(q)$ is the binary symmetric channel (that flips the measurement outcome) with error rate $q$.
		In general, $\mc P$ and $\mc N$ can represent arbitrary Pauli and binary channels.
	}
	\label{tab:ideal2simulated}
\end{table}

To simulate erasure circuits, we associate noise sources with each operation. We do this by adding erasure locations between every operation. Then, we append Pauli noise $\mc P$ after every operation and include bit-flip noise $\mc N$ on classical outcome bits (see Table~\ref{tab:ideal2simulated}). We use $e$ to denote erasure rates at erasure locations, $p$ to denote Pauli error rates, and $q$ to denote classical bit-flip rates. The noise strengths may vary depending on the operations around the error location. In a simplified model, we may consider Pauli channels to be depolarizing and $e$, $p$, and $q$ to be constant within the circuit.

At an erasure location, the qubit has a probability $e$ of being taken to an erasure subspace that is orthogonal to the computational subspace.
We assume that any computational subspace measurement involving an erased qubit results in a randomized outcome.
When an erased qubit interacts with another qubit through an entangling gate, we assume that a fully depolarizing channel is applied to the other qubit unless stated otherwise; see Sec.~\ref{Biased_noise_XZZX} for a physical model where erasure spreads via dephasing noise.\footnote{In the model where the other qubit is erased, erasure information may be obtained from both qubits, effectively increasing the confidence of the erasure checks, but at the cost of potentially more erasure spread between resets. Therefore, we expect the erasure-erasure spread mechanism to favor more frequent erasure checks.}
A reset operation reinitializes an erased qubit as the maximally mixed state in the computational subspace but acts trivially on qubits that are not erased. Alternatively, we may consider a unitary reset operation that exchanges the erased state with a fixed state in the computational basis. 
Reset operations may be conditioned on the classical EC outcomes.
We compare different reset schemes in Sec.~\ref{subsec:FNFP}.

In our formalism, no coherences are created between the computational and erasure subspaces.\footnote{We approximate the state after a unitary reset operation as an incoherent mixture.} This allows us to efficiently sample from erasure circuits. We use one bit of information per qubit to keep track of its erasure state and update it appropriately at erasure locations and reset operations. For qubits in the computational subspace, we use the Gottesman-Knill theorem~\cite{Gottesman1998,Aaronson2004} to represent the state and simulate the stabilizer circuit operations.

\subsection{Surface code as quantum memory}

\begin{figure}[htpb]
	\centering
	\raisebox{36mm}{(a)}\includegraphics[width=0.7\columnwidth]{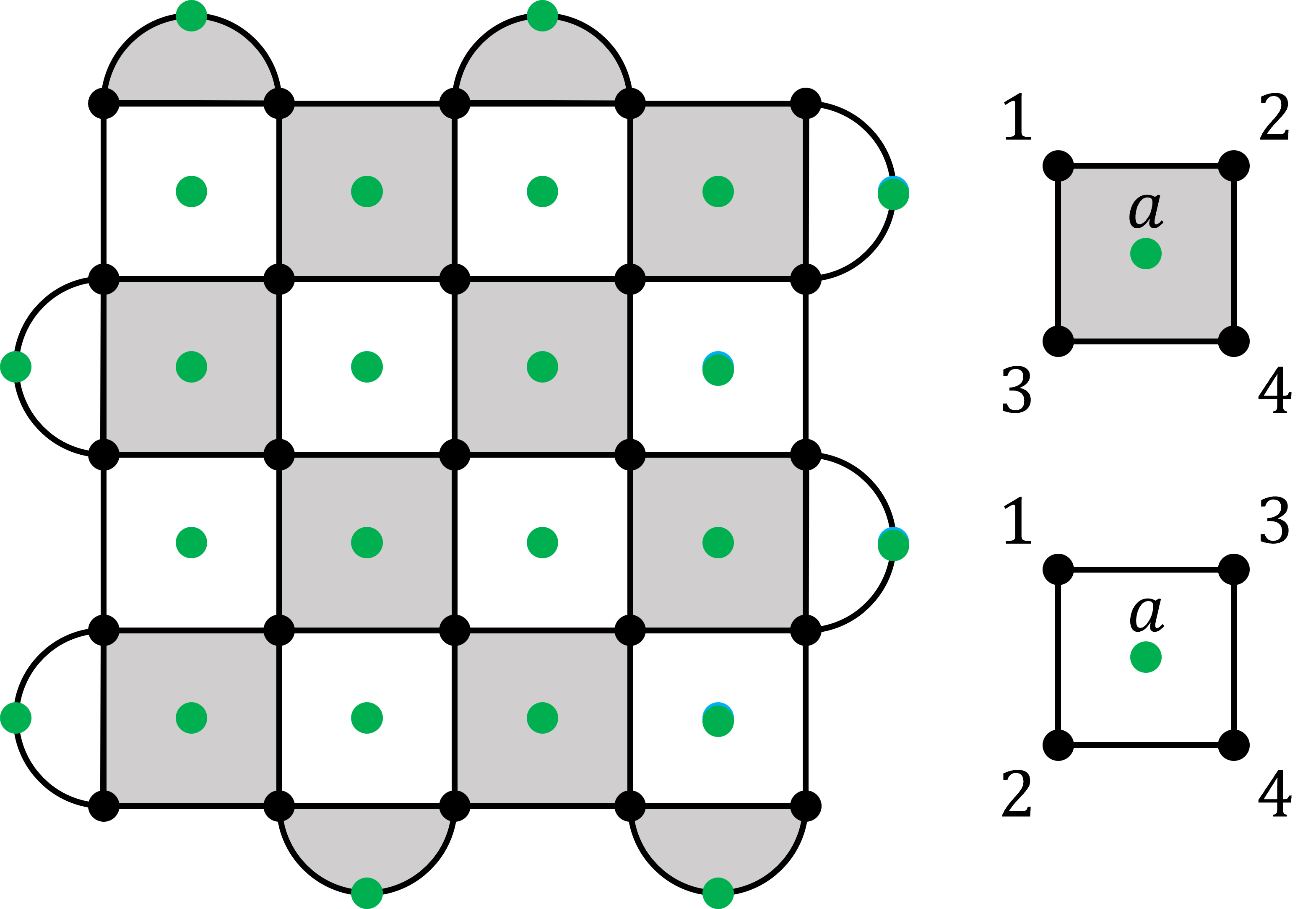}\\\vspace{1em}
	(b)\qquad$\Qcircuit @C=.7em @R=.7em {
		\lstick{1} & \qw & \qw & \qw & \gate{P} & \gate{A} & \qw & \gate{B} & \qw & \gate{C} & \qw & \gate{D} & \qw\\
		\lstick{2} & \qw & \qw & \qw & \qw & \gate{A} & \gate{P} & \gate{B} & \qw & \gate{C} & \qw & \gate{D} & \qw\\
		\lstick{3} & \qw & \qw & \qw & \qw & \gate{A} & \qw & \gate{B} & \gate{P} & \gate{C} & \qw & \gate{D} & \qw\\
		\lstick{4} & \qw & \qw & \qw & \qw & \gate{A} & \qw & \gate{B} & \qw & \gate{C} & \gate{P} & \gate{D} & \qw\\
		\lstick{a} & & & \lstick{\ket{+}} & \ctrl{-4} & \gate{A} & \ctrl{-3} & \gate{B} & \ctrl{-2} & \gate{C} & \ctrl{-1} & \meter_{\qquad\qquad\quad X}}$
	\caption{
		(a) The surface code with code distance $d=5$.
		The ancilla qubits (green dots) are used to measure stabilizer checks on the adjacent data qubits (black dots).
		Grey and white regions are associated with $X$ and $Z$ stabilizers, respectively.
		(b) Syndrome extraction circuit for the surface code, where the locations of possible ECs with reset are labeled by $A$, $B$, $C$, and $D$.
	}
	\label{fig_surface}
\end{figure}

The surface code is one of the most studied quantum error-correcting codes~\cite{Dennis2002,fowler2012surface,google2023suppressing}.
As a CSS stabilizer code, its codespace is defined as the +1 eigenspace of a set of commuting Pauli $X$ and $Z$ operators associated with the faces of a square lattice; 
see Fig.~\ref{fig_surface}(a).
To measure Pauli $X$ and $Z$ operators, we use extra ancilla qubits and implement standard syndrome extraction circuits, as depicted in Fig.~\ref{fig_surface}(b).
Importantly, the order of controlled-Pauli gates is chosen to minimize the error spread and to avoid hook errors~\cite{Dennis2002,tomita2014low}.
By repeating measurements of Pauli $X$ and $Z$ operators, we can learn reliable information about the errors afflicting the system.
Consequently, we can find an appropriate recovery operator and protect the encoded logical information from the errors.

Implementing the surface code with erasure qubits requires including ECs and reset operations in the syndrome extraction circuit.
A natural schedule, which we call the 1 EC schedule, is to perform ECs and reset operations on data qubits at locations $D$ in Fig.~\ref{fig_surface}(b) while ancillas are being measured.
Note that erasure information about the ancilla qubits will also be learned during the readout process according to Table~\ref{tab:ideal2simulated}, which is typical for many implementations of erasure qubits.
If ECs with reset take similar time as measurement and reinitialization of ancilla qubits, thus assuming operations are coherence limited this schedule will incur neither time nor noise penalty, while providing information about erasures.

However, performing only one EC per qubit during each syndrome extraction round may provide too little information about erasures, annulling the possible benefits of erasure qubits.
We therefore consider two additional schedules with more ECs and reset operations.
The 2 EC schedule has two ECs with reset per syndrome extraction round placed at locations $B$ and $D$ in Fig.~\ref{fig_surface}(b), and the 4 EC schedule places ECs with reset at all four locations $A$, $B$, $C$, and $D$.
Note that performing two or more ECs with reset per syndrome extraction round will lengthen the syndrome extraction round compared to the standard approach.

\section{Architecture-independent results}
\label{sec:generalresults}
In this section, we present our results on implementing the surface code with erasure qubits, without restricting ourselves to any particular architecture.
The simulations are performed by initializing an eigenstate of a logical Pauli operator for the surface code and sampling using the method outlined in Sec.~\ref{subsec:erasurequbitsoverview}.
At the end of the simulation, we decode by approximating the erasure circuit as a stabilizer circuit and applying minimum-weight perfect matching on the resulting decoding graph (see Sec.~\ref{subsec:approximatedecoding}).
An error is reported if after decoding, the value of the logical Pauli operator measured at the end of the circuit is different than the initial value.
Running the experiment many times for a distance-$d$ code gives us the logical error rate per $d$ syndrome extraction rounds, denoted $p_L$. For more simulation details, see Appendix~\ref{app:simulation_details}.

The simulations assume the simplified error model of Sec.~\ref{subsec:erasurequbitsoverview}, where the error parameters $e$, $p$, and $q$ are constant for the different operations within the circuit.
However, results about physical situations where noise parameters vary by operation may be inferred from the simulation data obtained in this section, for example, as done in Sec.~\ref{sec:superconducting}.

\subsection{Approximate solutions to the decoding problem}
\label{subsec:approximatedecoding}

Running a QEC protocol with erasure qubits requires correcting the errors that occurred using the available syndrome and EC outcomes.
We propose a variation of the decoding method for erasure circuits in Ref.~\cite{gu2023faulttolerant} that is suitable for schedules with infrequent ECs.

Erasure circuits can be decoded in three steps.
In the first step, we use the EC outcomes to map the erasure circuit to a stabilizer circuit with independent error mechanisms which are binary random variables.
This allows us to decode using standard techniques for stabilizer circuits.
In particular, we obtain a decoding hypergraph with error mechanisms as hyperedges and detectors as vertices.
By definition, detectors correspond to products of measurement outcomes that are deterministic in the absence of qubit and measurement errors. Each hyperedge in the decoding hypergraph consists of all of the detectors that would be flipped if the error mechanism occurs.
In the second step, we approximate the hypergraph as a graph by decomposing the hyperedges into edges, i.e., approximating error mechanisms with ones that cause at most two violated detectors.
In the third step, we use minimum-weight perfect matching on the graph from the previous step to find a likely error causing the syndrome.

Ref.~\cite{gu2023faulttolerant} described a method to do the first step exactly. Although the number of error mechanisms added in the conversion process is proportional to the size of the circuit if reset operations occur on every qubit at constant time intervals, that constant is exponential in the length of those intervals. For example, with four entangling gates between reset operations, the 1 EC schedule would introduce 1023 different error mechanisms per reset operation, making decoding impractical.

Here, we introduce a way to approximately convert erasure circuits to stabilizer circuits that results in fewer added error mechanisms. The idea is that the large number of error mechanisms is due to erasures causing correlated depolarization at many spacetime locations. Instead, we can approximate the converted stabilizer circuit by one with independent 1Q depolarizing errors, but still maintain the marginal probability of error at each spacetime location. The number of error mechanisms added to the resulting decoding hypergraph is only linear in the number of gates between reset operations.

We proceed as follows on an erasure circuit $C_E$. Consider a segment $s$ of a qubit $q$, which is defined to be the worldline of $q$ between two consecutive reset operations (see Fig.~\ref{fig:erasure2stabilizer}(a)). We will remove the erasure locations, ECs, and reset operations in $s$ and add appropriate 1Q depolarizing channels to approximate the effects of erasures in $s$. Doing this for all segments, we convert $C_E$ to a stabilizer circuit $C$.

\begin{figure}[ht]
	\centering
	(a) $\Qcircuit @C=.8em @R=.7em {
		&  & \lstick{\ldots} & \multigate{1}{G_1} & \rstick{\ldots} \qw\\
		\lstick{q} & \gate{\mathrm R} & \measure{\mc E} & \ghost{G_1} & \measure{\mc E} & \ghost{I} & \lstick{\ldots} & \multigate{1}{G_{r}} & \measure{\mc E} & \gate{\mathrm{R}} & \qw \\
		& & & & & & \lstick{\ldots} & \ghost{G_{r}} & \rstick{\ldots} \qw }$\\
	\vspace*{10pt}
	(b)\qquad\qquad $\Qcircuit @C=.8em @R=.4em {
		\lstick{\ldots} & \multigate{1}{G_1} & \measure{\mc F_1} & \rstick{\ldots} \qw\\
		\lstick{q} & \ghost{G_1} & \ghost{G} & \lstick{\ldots} & \multigate{1}{G_{r}} & \qw & \measure{\mc F_{r+1}} & \qw\\
		& & & \lstick{\ldots} & \ghost{G_{r}} & \measure{\mc F_{r}} & \rstick{\ldots} \qw }$
	\caption[A segment of an erasure circuit and the approximate conversion to a stabilizer circuit.]{
		(a) A segment $s$ of the qubit $q$ in an erasure circuit $C_E$.
		(b) The corresponding portion of the converted stabilizer circuit $C$. Depolarizing channels are placed at spacetime locations $\mc F_i$ according to Algorithm~\ref{alg:approxconversion}.}
	\label{fig:erasure2stabilizer}
\end{figure}

For $i\in\{1, \dots, r\}$, let $G_i$ be the $i$-th entangling gate in $s$ and $\mc F_i$ be the spacetime location placed after $G_i$ on the other qubit. For convenience, let $G_{r+1}$ denote the second reset operation in $s$ and $\mc F_{r+1}$ be the location at $q$ after the second reset operation. (See Fig.~\ref{fig:erasure2stabilizer}(b) for an illustration.) If $q$ was erased before $G_i$, the operation $G_i$ will result in full depolarization at spacetime location $\mc F_i$. Let $\bar a_i$ be the probability of this event, conditioned on the EC outcomes. The distribution of errors introduced by erasures in $s$ is approximated by independently sampling events with probabilities $\bar a_1, \dots \bar a_{r+1}$ and placing fully depolarizing channels at spacetime location $\mc F_i$ if the event with probability $\bar a_i$ is sampled. In other words, 1Q depolarizing channels of error probability $3\bar a_i/4$ are placed at locations $\mc F_i$. Note that each of these depolarizing channels is equivalent to three independent error mechanisms which apply each of the nontrivial Pauli operators $X$, $Y$ and $Z$ with probability $\frac 1 2 \left(1 - \sqrt{1 - \bar a_i}\right)$~\cite{Chao2020surfacecode}.

The approximate conversion of an erasure circuit to a stabilizer circuit with independent error mechanisms is summarized in Algorithm~\ref{alg:approxconversion}. See Appendix~\ref{app:decodingexample} for an example highlighting the difference between this method and the exact conversion of Ref.~\cite{gu2023faulttolerant}.

\begin{figure}[ht]
	\centering
	\raisebox{50mm}{(a)}\includegraphics[width=0.45\textwidth,trim={0 0 0 0},clip]{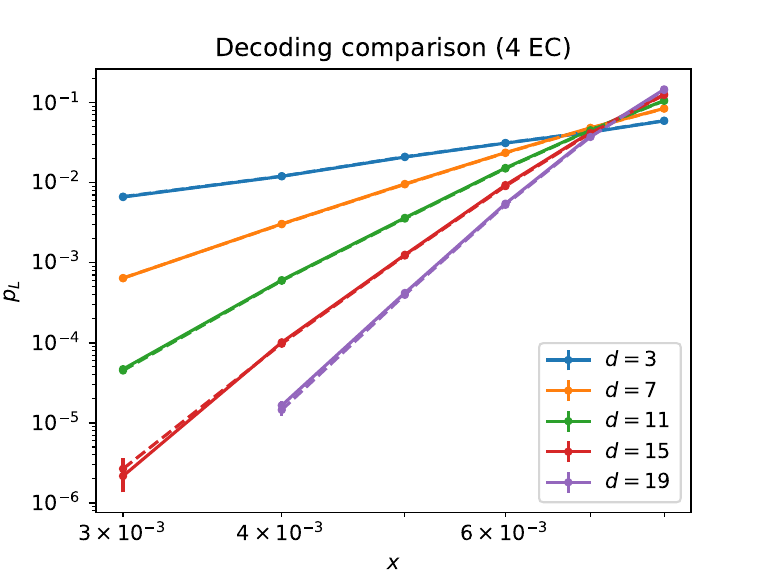}
	\raisebox{50mm}{(b)}\includegraphics[width=0.45\textwidth,trim={0 0 0 0},clip]{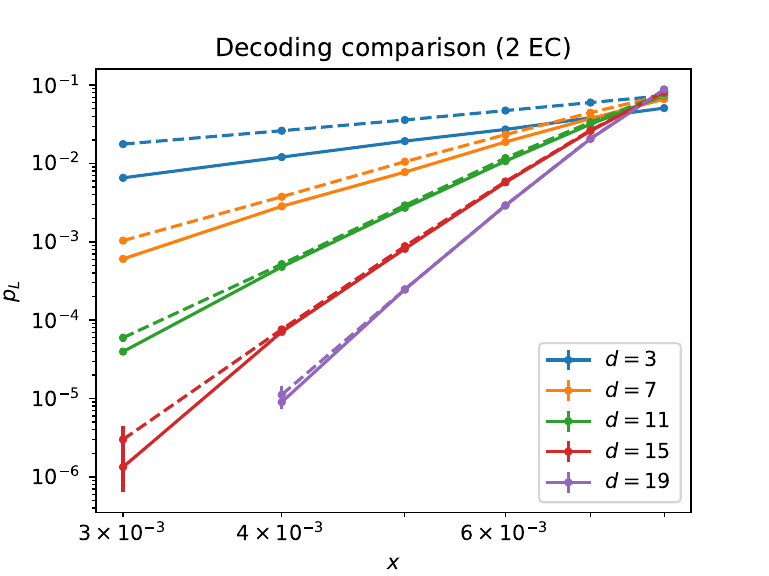}
	\caption{
		Comparison of the exact and approximate conversion schemes for decoding.
		We plot the performance for the (a) 4 EC and (b) 2 EC schedules.
		The solid lines correspond to approximate conversion of the erasure circuit to a stabilizer circuit;
		the dashed lines correspond to the exact method.
		The logical error rates are plotted for different values of $x$ along the line $(e, p, q) = (x, x/10, x)$.}
	\label{fig:decodingcomparison}
\end{figure}

We remark that it is not straightforward to compare decoding using the exact conversion of Ref.~\cite{gu2023faulttolerant} with using the approximate method in this section.
Although the first step is more accurate using exact conversion, the resulting hyperedges of the decoding hypergraph may be decomposed less optimally (by, for instance, Stim~\cite{gidney2021stim}) in step two of the procedure.
This hypothesis is explored in Fig.~\ref{fig:decodingcomparison}, where we highlight the difference in performance between the two decoding methods. In both cases, we decode using the three steps outlined in this section, the only difference being the exact conversion of Ref.~\cite{gu2023faulttolerant} or the approximate conversion outlined here in step one. We see that the latter method performs just as well or even better than the former. The argument that this is due to step two is strengthened by the more pronounced performance gap in the 2 EC schedule compared to the 4 EC schedule. Because sparser reset operations result in larger correlations, this could result in larger inaccuracies from suboptimally decomposing bigger hyperedges into edges. Although the inefficiency of the first method prevents us from performing a similar comparison for the 1 EC schedule, the results for the 4 EC and 2 EC schedules demonstrate the effectiveness of the approximate circuit conversion method. In what follows, we choose to decode using the approximate conversion because it achieves similar (or lower) logical error rates while becoming computationally more efficient.

\linespread{1.12}\selectfont
\begin{algorithm}[H]
	\caption{Approximate conversion of an erasure circuit to a stabilizer circuit with independent error mechanisms}
	\label{alg:approxconversion}
	\textbf{Input:}\\
	erasure circuit $C_E$, erasure check outcomes $\Vec{d}$\\
	\textbf{Output:} \\stabilizer circuit $C$, error mechanisms $\{(P_{i,j},p_i)\}$
	\begin{algorithmic}[1]
		\State $S \gets \{\text{segments in }C_E\}$
		\ForEach{$s\in S$}
		\State $\{G_i\} \gets $ entangling gates in $s$
		\State $\{\mc F_i\} \gets $ spacetime locations associated with $s$
		\ForEach{$i$}
		\State $\bar a_i \gets \Pr\left(q \text{ erased before } G_i \middle| \vec{d}\right)$
		\State $p_i \gets \frac 1 2 \left(1 - \sqrt{1 - \bar a_i}\right)$
		\ForEach{nontrivial Pauli error $P_{i,j}$ at $\mc F_i$}
		\State include error mechanism $(P_{i,j},p_i)$
		\EndForEach
		\EndForEach
		\EndForEach
		\State $C \gets \text{$C_E$ with deleted erasure checks and reset}$
		\State \Return $C$, $\{(P_{i,j},p_i)\}$
	\end{algorithmic}
\end{algorithm}
\linespread{1}\selectfont

\subsection{Probing the correctable region}

\begin{figure*}[ht!]
	\centering
	\includegraphics[width=0.32\textwidth,trim={1.6cm 1.3cm 2cm 1.1cm},clip]{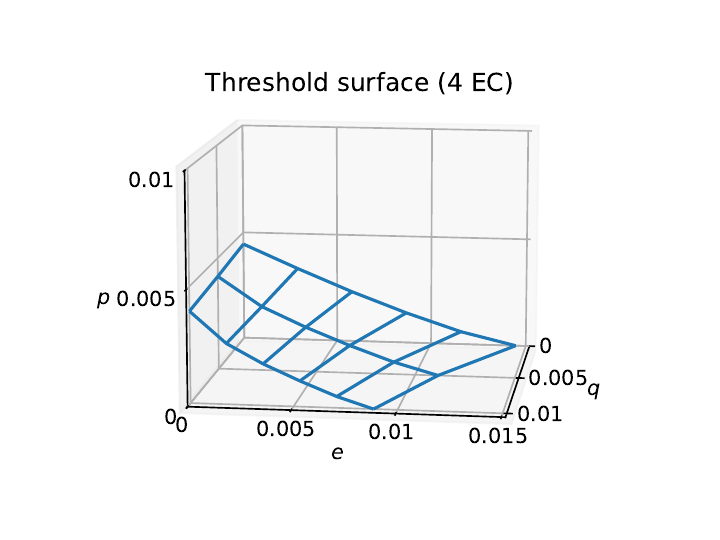}
	\hspace*{-0.32\textwidth}\raisebox{38.5mm}{(a)}\hspace*{0.295\textwidth}
	\includegraphics[width=0.32\textwidth,trim={1.6cm 1.3cm 2cm 1.1cm},clip]{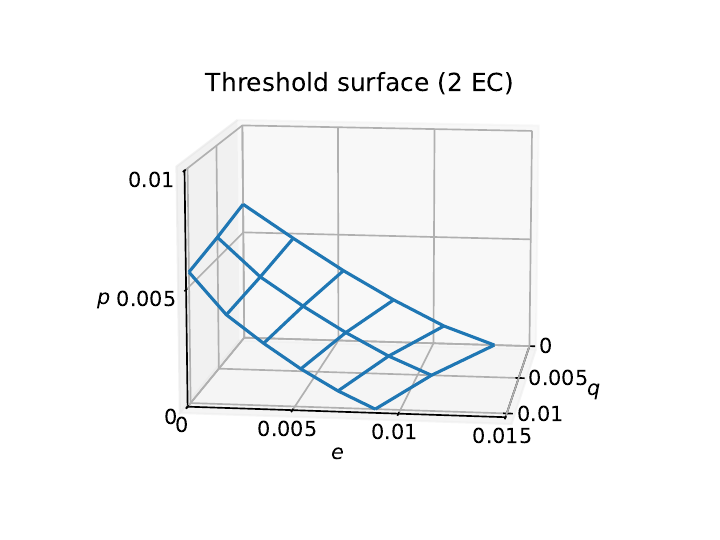}
	\hspace*{-0.32\textwidth}\raisebox{38.5mm}{(b)}\hspace*{0.295\textwidth}
	\includegraphics[width=0.32\textwidth,trim={1.6cm 1.3cm 2cm 1.1cm},clip]{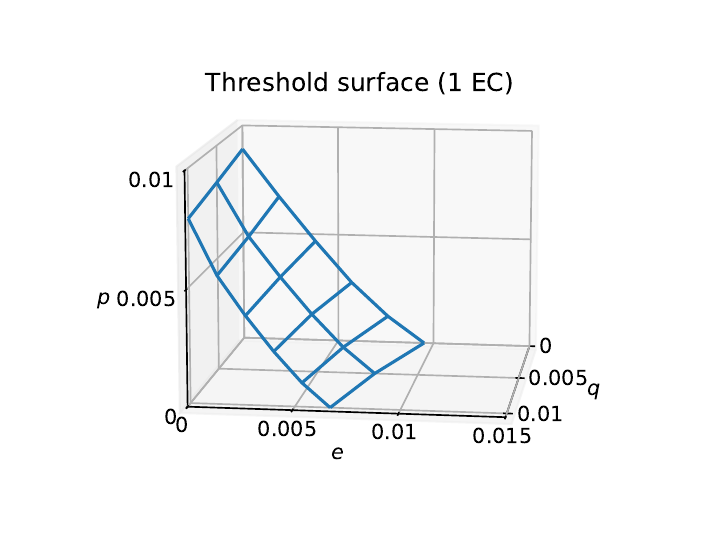}
	\hspace*{-0.32\textwidth}\raisebox{38.5mm}{(c)}\hspace*{0.295\textwidth}
	\newline
	\includegraphics[width=0.32\textwidth]{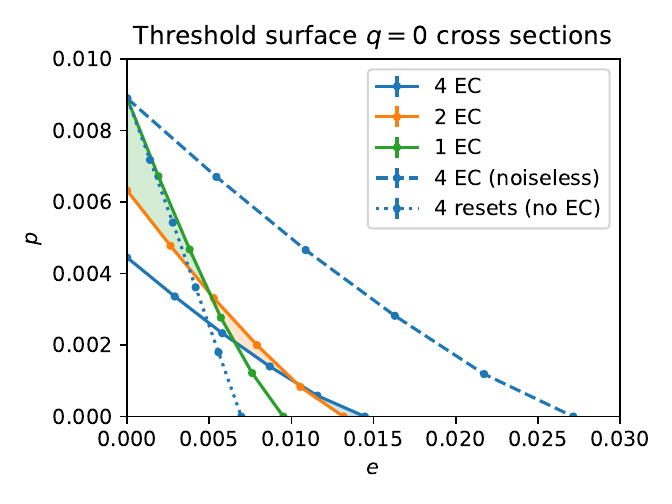}
	\hspace*{-0.32\textwidth}\raisebox{37mm}{(d)}\hspace*{0.295\textwidth}
	\includegraphics[width=0.32\textwidth]{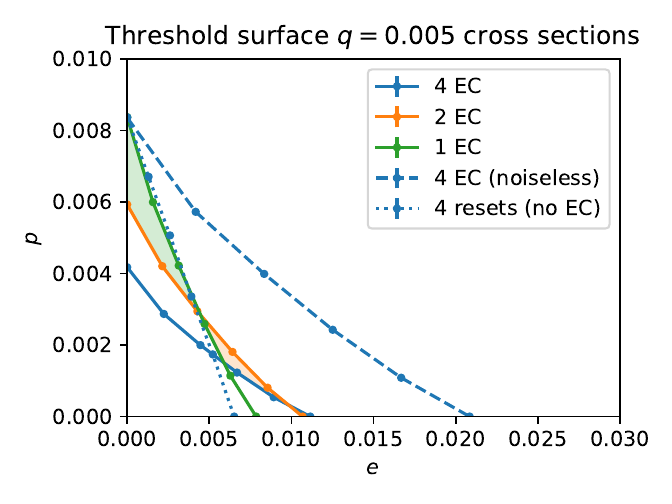}
	\hspace*{-0.32\textwidth}\raisebox{37mm}{(e)}\hspace*{0.295\textwidth}
	\includegraphics[width=0.32\textwidth]{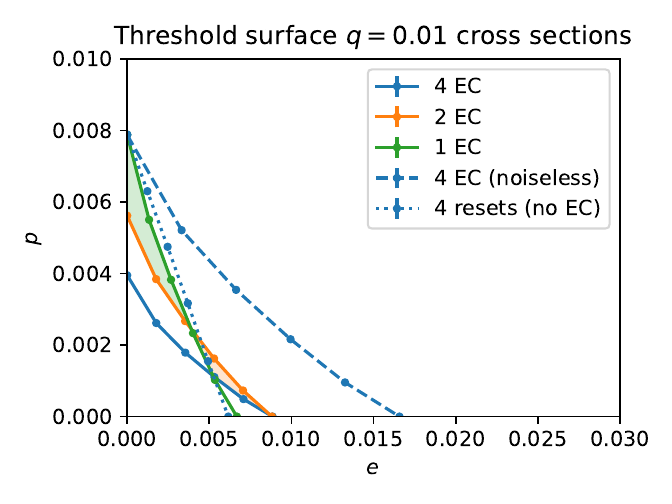}
	\hspace*{-0.32\textwidth}\raisebox{37mm}{(f)}\hspace*{0.295\textwidth}
	\caption{
		The correctable region in the $(e, p, q)$ phase space for the surface code with erasure qubits and EC schedules with: (a) 4 ECs, (b) 2 ECs, and (c) 1 EC, where $e$, $p$, and $q$ denote the erasure, Pauli, and measurement error rates, respectively.
		The meshes of the surfaces are aligned with constant values of $q$.
		Cross sections of the threshold surfaces in the $(e, p)$ plane for the different EC schedules when: (d) $q=0$, (e) $q=0.005$, and (f) $q=0.01$.
		The shaded regions indicate parameter regimes where only one schedule is below threshold.
		The dashed line corresponds to an optimistic scenario where ECs and reset operations do not cause additional errors.
		The dotted line is obtained from a scheme with noiseless reset operations but no ECs, giving an upper bound for schemes that do not use information about the locations of erasures.
	}
	\label{fig:thresholdsurfaces}
\end{figure*}

In Fig.~\ref{fig:thresholdsurfaces}(a)-(c), we plot the correctable region in the $(e, p, q)$ phase space for the different EC schedules.
The threshold surfaces bound the correctable regions where logical error rates are exponentially suppressed with increasing system size.

Fig.~\ref{fig:thresholdsurfaces}(d)-(f) shows cross sections of the threshold surfaces in the $(e, p)$ plane for various measurement error rates $q$, with the different curves representing different EC schedules.
As we increase the frequency of ECs, the $e$ threshold increases because we have more information about erasures, while the $p$ threshold decreases because there are more sources of noise.
However, because ECs themselves cause noise, the 4 EC schedule becomes less effective, even for high erasure rates, when the false detection rate $q$ is high.
We also plot the threshold curve (dashed line) for a scenario where the ECs with reset do not induce extra noise.
This optimistic situation arises if ECs are instantaneous or performed at the same time as entangling gates, assuming that noise is proportional to operation times.
The dotted line is an upper bound for standard QEC schemes that do not perform ECs.
Here, we assume that leaked qubits are periodically reset but that the reset operations do not introduce noise in the system.
The data indicates that the correctable region for erasure protocols almost entirely contains that of any standard scheme with free resets, and there is a significant regime of high erasure that is only correctable with erasure schemes.

It is important to note that defining EC as a separate circuit component results in different numerical values of the Pauli and erasure thresholds compared to earlier studies~\cite{kubica2023heralding}.
In those studies, the erasure rate is specified per CNOT gate without accounting for additional erasure mechanisms.
Here, since each qubit can be erased independently and may be erased during both entangling gates and ECs, there is a roughly $4\times$ difference between the reported values caused solely by different definitions.
For a more definition-independent analysis, we compare erasure and non-erasure qubits in Sec.~\ref{subsec:nonerasurecomparison}.

\subsection{Subthreshold performance}
Next, we explore the subthreshold scaling of the logical error rate. Knowing the functional form $p_L$ would allow us to estimate the minimal distance needed to achieve any given logical error rate for a set of physical parameters $(e, p, q)$. For low physical error rates, the probability of logical failure is dominated by the most likely error configurations~\cite{watson2014pLscaling}. In the phenomenological setting, a heuristic for the logical error rate under Pauli noise is $p_L \propto \left(p/p^*\right)^{\lceil d/2\rceil}$, where $p^*$ is the Pauli noise threshold. Similarly, pure erasure noise would give the scaling $p_L \propto \left(e/e^*\right)^{d}$, where $e^*$ is the erasure threshold.

\begin{figure*}[ht!]
	\centering
	\hspace{-1mm}\begin{minipage}{0.69\linewidth}
		\raisebox{33.5mm}{(a)}\includegraphics[width=0.4\textwidth,trim={1.6cm 1.3cm 2cm 1.1cm},clip]{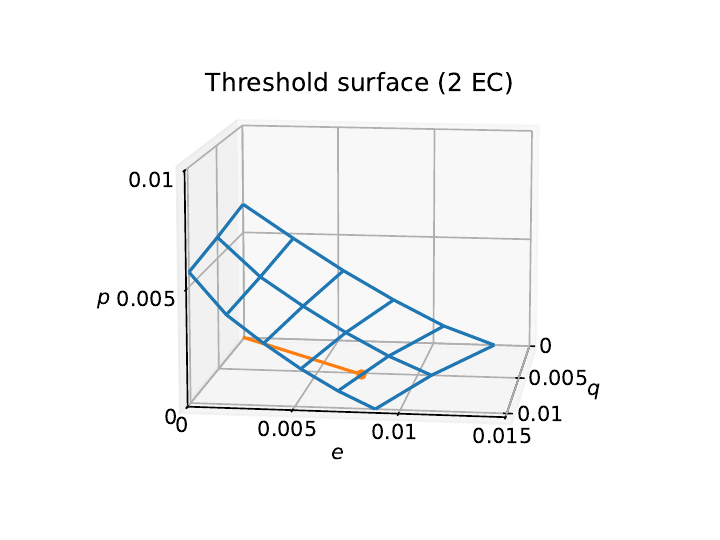}
		\raisebox{33.5mm}{(b)}\includegraphics[width=0.5\textwidth,trim={0 0 0 0},clip]{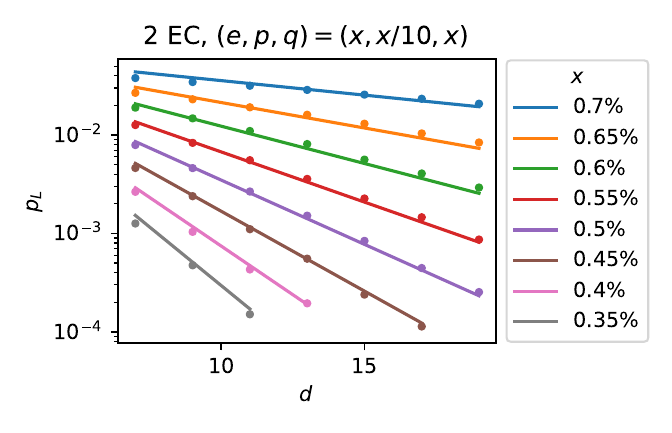}
	\end{minipage}
	\hspace{1mm}
	\begin{minipage}{0.28\linewidth}
		\raggedright (c)\\\vspace{2mm}
		\scriptsize
		\centering
		Values of $\alpha$ for different lines $\ell$\\
		\vspace{1mm}
		\begin{tabular}{| c | c | c | c |}
			\hline
			\boldmath$(t_e, t_p, t_q)$ & \textbf{4 EC} & \textbf{2 EC} & \textbf{1 EC} \\\hline
			$(1, 0.01, 1)$ & 0.84 & 0.76 & 0.64 \\\hline
			$(1, 0.1, 1)$ & 0.72 & 0.69 & 0.60 \\\hline
			$(1, 0.5, 1)$ & 0.61 & 0.59 & 0.57 \\\hline
			$(1, 5, 5)$ & 0.51 & 0.51 & 0.51 \\\hline
		\end{tabular}
		\vspace{7mm}
	\end{minipage}
	\caption{The subthreshold scaling of the logical error rate according to the ansatz of Eq.~\eqref{eq:subthreshold_ansatz}. (a) We test logical error rates along the orange line $\ell$ in the $(e, p, q)$ phase space.
		The orange dot is at the intersection of $\ell$ with the threshold surface, giving the threshold value $x^*$. (b) The logical error rates with increasing distance for various values of $x$ along $\ell$, shown in the different colors. To fit the data we use the numerical ansatz $p_L = a\left(x/x^*\right)^{\alpha d}$ with fitting parameters $a$, $\alpha$. (c) Values of $\alpha$ for various noise biases in the different EC schedules.
	}
	\label{fig:subthreshold_ansatz}
\end{figure*}

\begin{figure*}[ht!]
	\centering
	\includegraphics[width=0.32\textwidth,trim={0 0 0 0},clip]{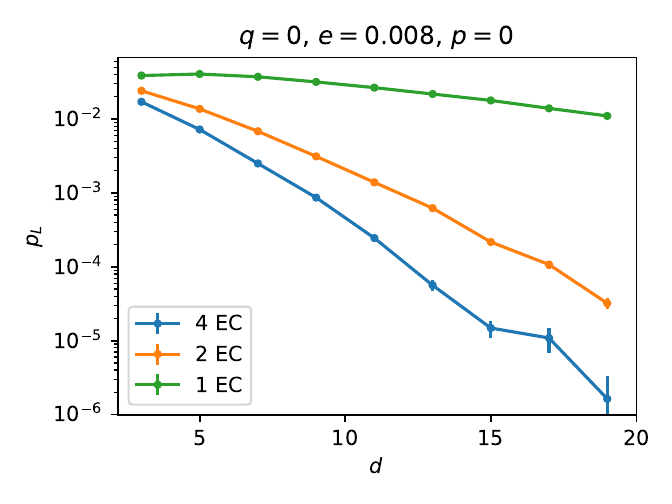}
	\hspace*{-0.32\textwidth}\raisebox{37mm}{(a)}\hspace*{0.295\textwidth}
	\includegraphics[width=0.32\textwidth,trim={0 0 0 0},clip]{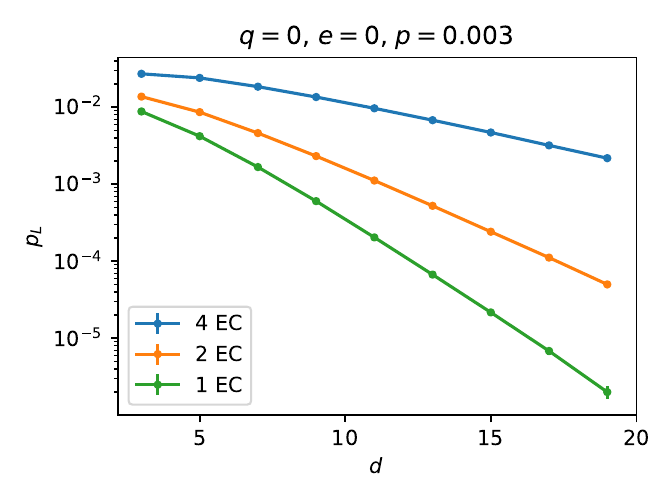}
	\hspace*{-0.32\textwidth}\raisebox{37mm}{(b)}\hspace*{0.295\textwidth}
	\includegraphics[width=0.32\textwidth,trim={0 0 0 0},clip]{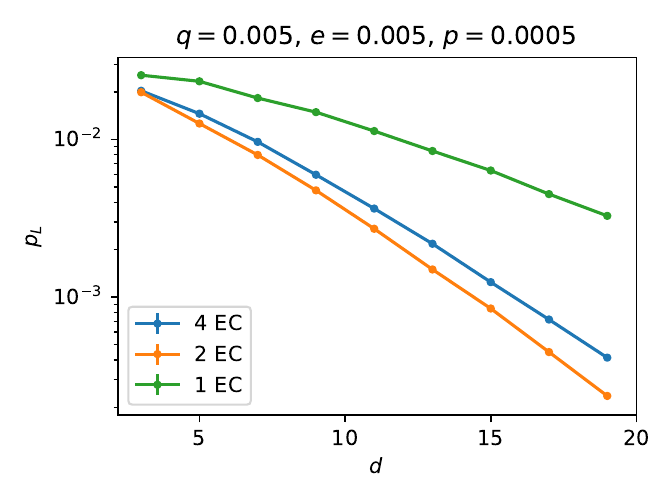}
	\hspace*{-0.32\textwidth}\raisebox{37mm}{(c)}\hspace*{0.295\textwidth}
	\caption{
		Scaling of the logical error rate with the distance $d$ for noise parameters $e$, $p$ and $q$ in the correctable region.
		(a) For pure erasure noise, the 4 EC schedule performs the best.
		(b) For pure Pauli noise, ECs introduce unnecessary noise, so the 1 EC schedule is the best. (c) For erasure biased noise, the 2 EC schedule is the best.}
	\label{fig:subthresholdscaling}
\end{figure*}

In our setting, we wish to determine the logical error rate for any point $(e, p, q)$ in the correctable region bounded by the threshold surface. We propose an ansatz of the following form.
Let $x$ be a single parameter characterizing a line $\ell$ through the origin $(e, p, q) = (t_e x, t_p x, t_q x)$ for constants $t_e, t_p, t_q$; see Fig.~\ref{fig:subthreshold_ansatz}(a).
The line $\ell$ describes different physical error rates of a given noise bias. Within the correctable region, we expect the logical error rate to scale along $\ell$ as
\begin{equation}
	\label{eq:subthreshold_ansatz}
	p_L \approx a\left(\frac{x}{x^*}\right)^{\alpha d}\, ,
\end{equation}
where $x^*$ is the value of $x$ at the intersection of $\ell$ with the threshold surface.
The quantities $a$, $\alpha$ are fitting parameters that depend on $\ell$. The parameter $a$ is the logical error rate at the threshold, which we find to be roughly constant for all lines $\ell$ and EC schedules. Importantly, $\alpha$ characterizes the effective distance of the scheme, interpolating between $1/2$ for pure Pauli noise and $1$ for erasure noise in the phenomenological setting.

To test how well Eq.~\eqref{eq:subthreshold_ansatz} describes the subthreshold logical error rate scaling, we plot $p_L$ against $d$ for different values of $x$ along a certain line $\ell$; see Fig.~\ref{fig:subthreshold_ansatz}(b).
The solid lines, indicating the expected logical error rates from the ansatz by fitting $a$ and $\alpha$, do approximate the data closely. Values of $\alpha$ for different noise biases are reported in the table Fig.~\ref{fig:subthreshold_ansatz}(c).
We see that $\alpha$ is close to $1/2$ when the erasure bias, defined as the ratio $e/p = t_e/t_p$, is low, as expected for unknown Pauli noise.
Its value increases with the erasure bias of $\ell$.
From a theoretical standpoint, in the infinite erasure bias limit and in the absence of measurement errors, $\alpha$ should approach 1 for the 4 EC schedule because a single erasure cannot spread to Pauli errors on more than one data qubit. In contrast, $\alpha$ is lower for the less frequent EC schedules which allow more spread of erasures. The plots showing how well the ansatz in Eq.~\eqref{eq:subthreshold_ansatz} describes the data used to obtain the $\alpha$ values in the table are presented in Fig.~\ref{fig:supplementary_subthreshold_ansatz} of Appendix~\ref{app:simulation_details}.

In Fig.~\ref{fig:subthresholdscaling}, we compare the subthreshold scaling of $p_L$ with distance for the different EC schedules. We choose three sets of error parameters, each below threshold for all three EC schedules, that illustrate when each schedule should be used. For noise with very large erasure bias, Fig.~\ref{fig:subthresholdscaling}(a) illustrates that using four ECs per round suppresses logical error rates the best. When the noise is heavily biased toward Pauli errors, ECs are less useful, so the 1 EC schedule with the free EC is optimal (Fig.~\ref{fig:subthresholdscaling}(b)).
For a realistic scenario with a $10\times$ erasure bias and nonzero measurement error, 2 EC gives the best scaling (Fig.~\ref{fig:subthresholdscaling}(c)).

\section{Superconducting architectures}
\label{sec:superconducting}

A promising realization of an erasure qubit is via the dual-rail qubit \cite{Kubica2023,Levine2023}.
The dual-rail qubit converts energy relaxation, which is a fundamental noise mechanism in superconducting architectures, to erasures.
It is also robust to changes in the constituent transmons' energy gap, which significantly reduces the rate of Pauli $Z$ errors in its computational basis.
We now discuss how the physical properties of the dual-rail qubit translate to its error-correcting abilities.

\subsection{Dual-rail qubits}
\label{subsec:dual-rail_overview}
A dual-rail qubit is an erasure qubit composed of two coupled transmons. The computational subspace is spanned by basis states
\begin{equation}
	\ket{\bar b} = \frac{1}{\sqrt 2}\left(\ket{01} + (-1)^b\ket{10}\right), \quad b = 0, 1\, .
\end{equation}
The dominant noise mechanism for superconducting qubits is amplitude damping, where the excited state $\ket 1$ of one of the component transmons relaxes to the ground state $\ket 0$. This error removes the dual-rail qubit from the computational subspace, and the resulting orthogonal state $\ket{00}$ can be subsequently detected.

The dual-rail qubit is made up of two coupled resonant transmons. In the rotating wave approximation, they are described by the Hamiltonian
\begin{equation}
	H_0=\sum_{i=1,2}\left(\omega_{\text{DR}} a_{i}^{\dagger}a_{i}+\frac{\eta}{2}a_{i}^{\dagger}a_{i}^{\dagger}a_{i}a_{i}\right)+g_{12}\left(a_{1}^{\dagger}a_{2}+\text{h.c.}\right).
\end{equation}
Here, $i=1,2$ labels the transmons, $a_{i}$, $a_i^\dagger$ are the ladder operators, $\omega_{\text{DR}}$ and $\eta < 0$ are the transmon frequency and nonlinearity, respectively, and $g_{12}$ is the coupling frequency.
The computational basis states $\ket{\bar b}$ are the eigenstates of the dual-rail Hamiltonian with a single excitation and energies $ E_{\ket{\bar{b}}} =\omega_{\text{DR}} + (-1)^b g_{12}$.

The amplitude damping channel describes the process by which an excitation in the transmon escapes to the cold environment. It can be modeled by the jump operators $L_{i}=\sqrt{1/T_{1}^{\left(i\right)}}a_{i}$,
where $T_{1}^{\left(i\right)}$ denotes the energy relaxation time of transmon $i=1,2$.
Currently, typical energy relaxation rates for a high-quality transmon are $T_1\approx100\unit{\mu s}$~\cite{carroll2022dynamics}.
Within the computational subspace of the dual-rail qubit, the dephasing time is much slower than that of the constituent transmons and can be as high as $T^{(\text{DR})}_\phi=1 \unit{ms}$ since the dual-rail internal coupling behaves analogously to dynamical decoupling~\cite{Kubica2023}, as was demonstrated in Ref.~\cite{Levine2023}. ECs on dual-rail qubits can be implemented as described in Ref.~\cite{Kubica2023}.

One straightforward method to restore a dual-rail qubit from the erased state to the computational subspace is by using an XY drive, which we call a unitary pulse.
One can drive the qubit from the $\ket{00}$ state to a particular state within the computational subspace, e.g., the state $\ket{\bar{0}}$.
As the XY drive is a unitary operation limited to the transmon subspace, it will inevitably transform a computational state to the erased state $\ket{00}$\footnote{It is possible to design a pulse that transitions a computational state to a different non-computational state, but that would be even more detrimental.}.
As a result, this process should only be performed conditioned on detecting erasures, with false-positive erasure detection introducing additional erasures.

An alternative method of implementing the reset operation is by removing the uncertainty in the erasure status of the dual-rail qubit to a different degree of freedom, such as the readout resonator, and applying a single pulse that maps the erased state back to the computational subspace without dephasing the computational subspace.
We refer to such pulse as a one-way pulse.
Similar pulses were used for conditional reset and leakage reduction using the transmon's readout resonator~\cite{zhou2021rapid,magnard2018fast,geerlings2013demonstrating,marques2023all}.

Labeling the total state as $\ket{\text{dual-rail}}\otimes\ket{\text{resonator}}$, we can implement a pulse that transforms $\ket{00}\otimes\ket{0} \mapsto \ket{\bar 0}\otimes\ket{1}$ without disturbing the computational subspace. One way to achieve this is to use a high-frequency parametric drive. If the readout resonator is coupled symmetrically to the dual-rail qubit, the combined Hamiltonian is
\begin{align}
	H & =H_{0}+\delta\cos\left(\omega_{d}t\right)\sum a_{i}^{\dagger}a_{i}\nonumber\\
	& +\omega_{RO}r^{\dagger}r+g_{RO}\left(r^{\dagger}+r\right)\sum_{i=1,2}\left(a_{i}^{\dagger}+a_{i}\right)\, ,
\end{align}
where $r$, $r^\dagger$ are the ladder operators of the readout resonator, $\omega_{RO}$ is its frequency, and $g_{RO}$ is its coupling to the transmons. The frequency of both transmons are modulated with amplitude $\delta$ and frequency $\omega_d$. 

When $\omega_d= \omega_{DR} + \omega_{RO} + g_{12}$, the transition $\ket{00}\otimes\ket{0} \mapsto \ket{\bar 0}\otimes\ket{1}$ becomes resonant with Rabi frequency $g_{RO}\delta/\sqrt{2}\omega_{d}$. The next closest resonance is the transition from $\ket{\bar 1}\otimes \ket{0}$ to the state $\ket{B_+}\otimes\ket{1}$, where $\ket{B_+}$ is the closest state in the double excitation subspace to the $\ket{11}$ state (see Appendix~\ref{one_way_pulse_derivation}). The transition frequency between the two states is protected by the shift
\begin{equation}
	\Delta=\frac{\eta}{2}+\sqrt{\frac{\eta^{2}}{4}+4g_{12}^{2}}\,,
\end{equation} 
which in the $g_{12}\ll |\eta|$ regime is approximately ${4g_{12}^{2}}/{\left|\eta\right|}$ (since $\eta < 0$).

Standard transmons and cavities would require a very high frequency ($\gtrsim 10 \unit{GHz}$) drive. This can be partially mitigated by driving the transmon's frequency using a flux drive around its minimum or maximum frequency. At this point, the flux is only second-order sensitive to flux~\cite{didier2019ac}, which doubles the effective drive frequency at the expense of reducing the speed of the pulse.

It is also possible to implement the reset operation for the dual-rail qubit using quantum operations that preserve the erasure information.
For example, similar to Ref.~\cite{Levine2023}, the erasure detection can be implemented using an ancilla qubit. The ancilla is coupled to the dual-rail qubit, shifting its energy gap when the dual-rail qubit decays outside the computational subspace.
An XY drive at the shifted frequency excites the ancilla only when the dual-rail qubit is erased, without dephasing the computational subspace.
A reset pulse can be applied to the dual-rail qubit by applying a cross-resonance drive on the ancilla~\cite{magesan2020effective}, which would only excite the dual-rail qubit when the ancilla is excited.
The erasure information can later be recovered by measuring the ancilla. 

\subsection{Comparison with conventional transmons}
\label{subsec:nonerasurecomparison}
In Sec.~\ref{sec:generalresults}, we characterized the performance of different EC schedules in terms of error parameters $e$, $p$, and $q$. These general results can be used to draw conclusions about specific physical scenarios without obtaining additional data from simulations. To do this, we express the results in terms of physical parameters. For superconducting qubits, the relevant quantities are the amplitude damping $T_1$ and dephasing $T_\phi$ times, as well as the operation times $T_{2Q}$ for 2Q gates and $T_M$ for readout or ECs with reset. The goal in this section is to identify the conditions under which the performance of the dual-rail qubit exceeds that of the transmon and when it is advantageous to perform more frequent ECs.

In the dual-rail qubit, errors mainly arise from amplitude damping (causing erasures) and dephasing (causing Pauli errors). After time $t\ll T_1$, the probability of erasure is approximately $e = t/T_1$. Averaging over the syndrome extraction circuit, for the $l$ EC schedule, where $l=1,2,4$, we can estimate the effective erasure rate per operation as
\begin{equation}
	\label{erasure_def}
	e=\frac{4T_{2Q}+lT_{M}}{\left(4+l\right)T_{1}}\, .
\end{equation}
The probability of Pauli error is determined by the dephasing rate $T_\phi$. The parameter $p$ represents the probability of a gate failure and, for a 2Q gate, is equal to the probability that either of the qubits involved experiences an error. Averaging over the syndrome extraction circuit again, we obtain
\begin{equation}
	\label{eq:p_physical}
	p = \frac{4T_{2Q} + lT_M}{(2 + l)2T_\phi}\, ,
\end{equation}
where the factor of $2+l$ in the denominator is due to each 2Q gate contributing half an error location on the qubit.

When considering a single transmon, we consider an optimistic situation with no leakage. Hence, we approximate the scenario with $e=0$ and
\begin{equation}
	\label{eq:p_physical_transmon}
	p = \frac{4T_{2Q} + T_M}{(2 + 1)2T_\phi} + \frac{4T_{2Q}+T_{M}}{\left(2+1\right)2T_{1}}\, ,
\end{equation}
where we take $l=1$ in Eq.~\eqref{eq:p_physical} and add an additional noise mechanism with rate $1/(2T_1)$ modeling the qubit energy relaxation.\footnote{The factor of 2 can be thought of as the transmon occupying the ground state half the time, while the dual-rail qubit is always excited, so it decays faster.} We motivate these approximations in Appendix~\ref{appx:error_channels}.

\begin{figure}[ht!]
	\centering
	\raisebox{49mm}{(a)}\includegraphics[width=0.43\textwidth,trim={0 0 0 0},clip]{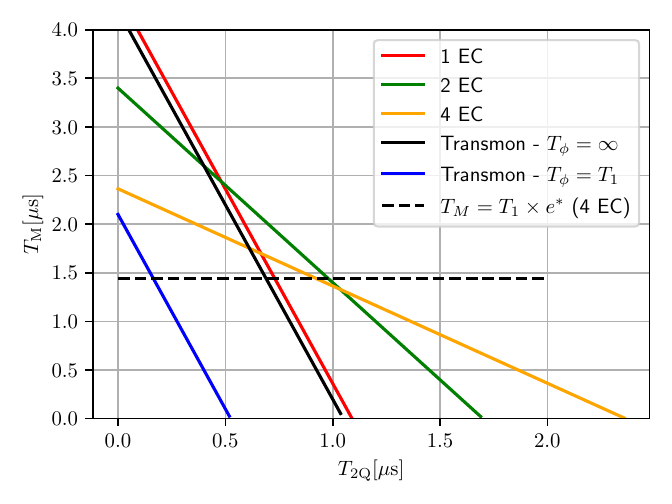}
	\raisebox{51mm}{(b)}\includegraphics[width=0.45\textwidth,trim={0 0 0 0},clip]{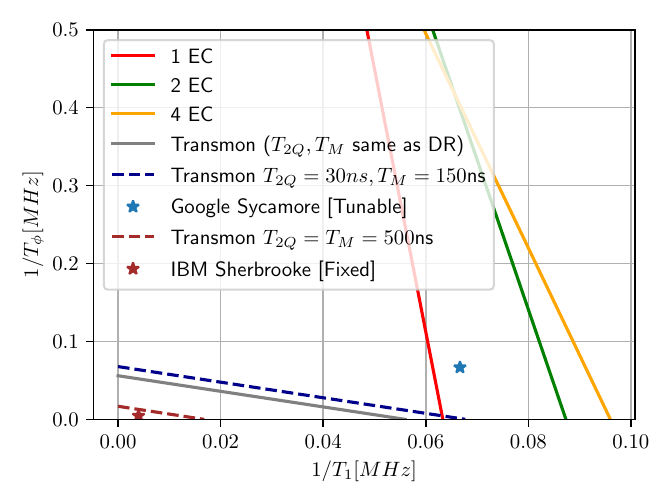}
	\caption{
		Cross sections of the threshold surfaces in the (a) $(T_{2Q}, T_M)$ and (b) $(T_1, T_\phi)$  planes for different EC schedules.
		In (a), we set $T_1=100\unit{\mu s}$ for both the dual-rail qubit and the transmon.
		The dual-rail qubit dephasing time is extended to $T^{(\text{DR})}_\phi=1\unit{ms}$ due to its internal coupling.
		For the transmon, we consider two dephasing times $T_\phi = 100\unit{\mu s},\infty$.
		The crossover regime between the 4 EC, 2 EC and 1 EC schedules occurs around $T_M=T_1\times e^*_{4\textrm{EC}}$.
		In (b), we set $T_{2Q}=T_{M}=150\unit{ns}$.
		The extended dephasing time of the dual-rail qubit is accounted for by setting $T_\phi^{(DR)} = 30 T_\phi$. To compare it with the transmon, we choose two regimes that approximate the gate times in Refs.~\cite{google2023suppressing,ibm_sherbrooke}. 
		Transmons with long $T_\phi$ are outperformed by $30$-$40\%$ in terms of minimal $T_1$, depending on how fast they implement 2Q gates. 
	}
	\label{fig:dual_rail_threshold}
\end{figure}

We now express the error parameters $e$, $p$ and $q$ that describe the threshold surfaces of the correctable regions in Fig.~\ref{fig:thresholdsurfaces} in terms of the physical parameters $T_1$, $T_\phi$, $T_{2Q}$ and $T_M$ of the transmon.
For simplicity, we assume $q=0$ and use the ansatz
\begin{equation}
	e/e^* + p/p^* = 1\, ,
\end{equation}
where $e^*$ and $p^*$ are the erasure and Pauli thresholds, to interpolate the cross section of the $(e,p,q)$ threshold surface.
We present the results in Fig.~\ref{fig:dual_rail_threshold}.

When we fix $T_1, T_\phi$, the dual-rail qubit can tolerate slower gates compared to the transmon.
Fig.~\ref{fig:dual_rail_threshold}(a) can be interpreted as follows.
When the ratio $\frac{T_\phi}{T_1}$ exceeds $\frac{e^*}{p^*}$, erasures become dominant.
The dual-rail qubit threshold curves are essentially those described by Eq.~\eqref{erasure_def}, with a threshold corresponding to $e^*$ (which depends on the EC schedule).
For the transmon curve where $T_\phi=\infty$, it aligns with Eq.~\eqref{eq:p_physical_transmon} with $p^*=0.007$ for a standard surface code.
The transmon curve with $T_\phi=T_1$ incorporates double the Pauli error.
In the limit of $T_\phi \to \infty$ (Fig.~\ref{fig:dual_rail_threshold}(a)), the maximal measurement time for the dual-rail qubit is similar to that of the transmon.
We see a transition regime where fewer detections become preferable to 4 EC, approximately at the point $T_M=T_1 \times e^*_{4\textrm{EC}}$, where $e^*_{4\textrm{EC}}$ is the erasure threshold value for 4 EC.

In addition to the threshold, we also consider optimizing with respect to the final logical error rate. The highest logical fidelity is achieved by optimizing over the scaling of the logical error rate in the distance, i.e., $\left(x/x^*\right)^\alpha$ as defined in Eq.~\eqref{eq:subthreshold_ansatz}. We plot the optimal protocol in various regions of parameter space in Fig.~\ref{fig:dual_rail_threshold_heatmap} of Appendix~\ref{choice_of_optimal_protocol}. As long as the gate and measurement times for the dual-rail qubit are not more than twice as long as those for transmons, the dual-rail qubit consistently outperforms the non-erasure scheme. A heuristic is that the optimal erasure detection protocol is typically the one corresponding to the point in 
$T_g/T_M$ space that lies farthest from the threshold along the line extending from the origin.

One would expect the dual-rail qubit to be advantageous compared to the transmon with the same $T_1$.
However, since the decay rate of the dual-rail qubit is twice the decay rate of the transmon and dual-rail gates tend to be slower, this benefit will only be significant if measurements for the dual-rail qubit can be realized on a par with measurements for the transmon.
At that point, the elevated threshold of the dual-rail qubit substantially reduces the minimal necessary $T_1$ to approximately half that of the transmon.

When we consider the the 2Q gate and measurement times $T_{2Q}$ and $T_M$, dual-rail qubits greatly outperform transmons with bad $T_\phi$ (Fig.~\ref{fig:dual_rail_threshold}(b)).
However, even with $T_\phi\to \infty$, as long as the dual-rail qubit can be measured at a similar rate, e.g., in $150\unit{ns}$, then the additional erasure information results in the minimal necessary $T_1$ reduced by $2\times$.

For transmons, there is a trade-off between tunability and gate time. Tunable transmons admit fast gates but worst coherence. Fixed transmons have much better coherence at the expense of slower gates. The dual-rail qubit has the advantage of both. Namely, it has a better threshold when gates are implemented slowly, and it can be tunable with a very long lifetime. In Fig.~\ref{fig:dual_rail_threshold}(b), we plot the parameters for two devices that roughly characterize the state-of-the-art for tunable~\cite{google2023suppressing} and fixed~\cite{ibm_sherbrooke} transmons.
The low $T_\phi$ of tunable transmons significantly inhibits their fidelity, greatly reducing the advantage of their fast gates.
Given that $T_1\gtrsim T_\phi$ for fixed transmons, the dual-rail qubit's robustness to dephasing offers a significant advantage.

More generally, state-of-the-art decoherence rates compared with gate and measurement times would suggest that both transmons and dual-rail qubits have long surpassed the QEC threshold, whereas this is a result that was only achieved recently~\cite{google2023suppressing,andersen2020repeated,google2025quantum}.
For transmons, this discrepancy is explained by both leakage to higher excited states and the difficulty to consistently fabricate many transmons with long coherence times on a single device.
Dual-rail qubits are more robust to fabrication issues since they have a wide tunability range with long coherence~\cite{Levine2023}. However, as demonstrated by the transmon scenario, coherence is not always the primary factor affecting fidelity, which could diminish the benefits of the dual-rail qubit.

Dual-rail qubits are more resistant to leakage during 2Q gates.
The energy gap in the computational subspace is approximately $20\times$ smaller than the gap to the next excited state, highly suppressing any leakage due to driven 1Q and 2Q gates. 
2Q gates preserve the number of excitations if the drive frequency is much lower than $\omega_{DR}$, but even such drives can cause leakage by transferring an excitation from a dual-rail qubit to its partner qubit.
This type of leakage is protected only by the detuning between the two dual-rail qubits and the transmon nonlinearity, but it erases the partner dual-rail qubit, making it easier to identify.
It is also reasonable to assume that leakage detection can be integrated into the erasure detection scheme.
Moreover, the inherent intrinsic noise cancellation opens up new avenues for developing dual-rail qubits from components that exhibit extended $T_1$ yet decreased $T_2$ coherence times. Candidates might include transmon-inspired circuits operating at lower $E_J/E_C$ ratios, phase qubits, or possibly fluxonium. Although these physical qubits, when used individually, might seem suboptimal due to their limited coherence, they could be transformed into exceptionally robust logical qubits in a dual-rail configuration, significantly extending erasure times while maintaining strong phase coherence within the dual-rail subspace.

\subsection{Imperfect erasure reset}
\label{subsec:FNFP}
Next, we examine how errors in erasure detection impact the performance of dual-rail qubits. To clarify, we define false-negative erasure detections as instances where actual erasures are missed, and false-positive detections as cases where erasures are incorrectly identified.

We also compare the two implementations of reset operations and find their effect on the QEC threshold.
We found that implementing one-way pulses following the erasure detection yields the best overall performance. While both false-positive and false-negative erasure detection reduced the erasure threshold, this reduction was approximately halved when using the one-way pulse after the detection.

To simplify the analysis, we simulated noise that only occurs during gates~\cite{Kubica2023} and expressed the results as the averaged erasure per operation. This modification slightly shifts the value of $e^*$, but does not change the qualitative results.

The process of restoring the dual-rail qubit may need to incorporate the classical feedback and operate on a time scale similar to that of other quantum operations.
Here, we show that the reset operation of the dual-rail qubit that does not depend on the erasure information amplifies the noise from erroneous erasure detections.
This type of classical feedback has been demonstrated in superconducting qubits \cite{corcoles2021exploiting,stefanazzi2022qick}, with an extra delay that, in principle, can go below $50\unit{ns}$.
Single Flux Quantum (SFQ) devices~\cite{mukhanov2019scalable} have the potential to further decrease feedback latency to a minimal level.

While fast feedback is achievable and has been demonstrated in leading platforms, many current devices are still limited by feedback delays due to hardware or integration constraints. The ancilla-based recovery pulse, described in Sec.~\ref{subsec:dual-rail_overview}, that preserves the erasure information can allow some additional leeway in designing the cycle of current devices. This can be combined with a decreased frequency of ECs, reducing the effect of the measurement time on the erasure rate in Eq.~\eqref{erasure_def} (albeit also lowering the QEC threshold).

In systems with a high rate of false-negative erasure detections, one might consider using the one-way pulse on all of the qubits following every EC, independently of the EC outcome.
Erasure qubits that were missed due to a false-negative detection would return to the computational subspace without raising any erasure flag.
This approach would result in the loss of information regarding the qubit decay but would prevent erasures from spreading and harming additional qubits.
However, we find through numerical simulations that the information about erasures is more valuable, and pulsing all the qubits effectively reduces the QEC threshold.

\begin{figure}[htpb]
	\raisebox{52mm}{(a)}\includegraphics[width=0.45\textwidth]{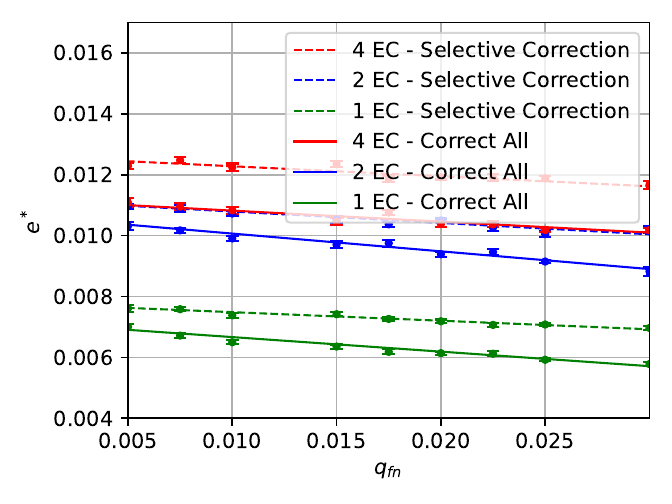}
	\raisebox{51mm}{(b)}\includegraphics[width=0.45\textwidth]{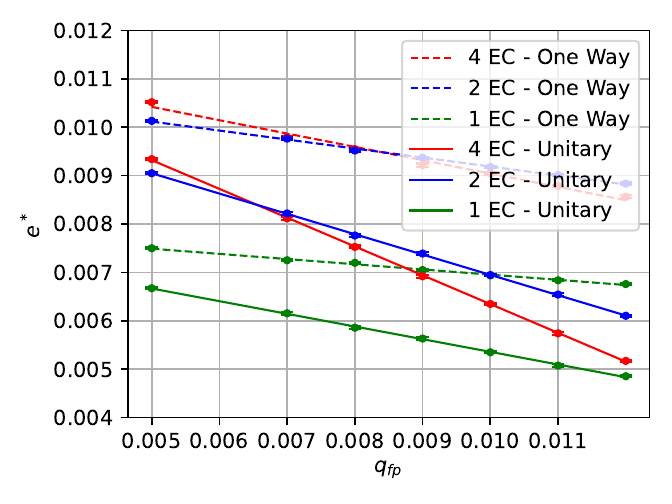}
	\caption{
		Cross sections of the threshold surfaces in the (a) $(e, q_\text{fn})$ and (b) $(e, q_\text{fp})$ planes, where $q_\text{fp}$ and $q_\text{fn}$ are the rates of false-positive and false-negative erasure detections, respectively.
		In (a), we use either a selective reset pulse on qubits that were identified as erased, or apply a reset pulse to all qubits.
		The trend lines are fitted using Eq.~\eqref{eq:fp_threshold}.
		In (b), we use the one-way or unitary pulse to reset erased qubits.
		The trend lines fitted using Eq.~\eqref{eq:false_positive_scaling}.
		\label{fig:false_negative} Error bars represent a single standard deviation. 
	}
\end{figure}

Assuming no Pauli errors, i.e., $p=0$, we find the maximal erasure rate $e$ for a range of false-negative erasure detection rates $q_\text{fn}$; see Fig.~\ref{fig:false_negative}(a).
We note that the data can be fitted with the ansatz
\begin{equation}
	\label{eq:fp_threshold}
	\frac{e}{e^{*}}+\frac{q_{\text{fn}}}{q_{\text{fn}}^{*}}=1\, ,
\end{equation}
where $q^*_{\text{fn}}$ is a fitting parameter.
When the reset pulse is applied selectively to the qubits once they are identified as erased, we obtain $q_\text{fn}^* \approx 0.49, 0.33, 0.28$ for the 4 EC, 2 EC and 1 EC schedules, respectively.
When the reset pulse is applied to all qubits following every EC, we find $q_\text{fn}^* \approx 0.3, 0.17, 0.14$. The standard deviation of our fit was less than $1\%$ of each of these values.

We implemented numerical simulations to assess the impact of false-positive erasure detection on the logical error rate.
When a false-positive detection occurs with a one-way pulse, the qubit is not erased in the simulation, but the decoder is (incorrectly) informed about an erasure.
The decoder knows the probability of false-positive detections and adjusts the probabilities on the matching graph accordingly. 

To simulate the unitary pulse, following each erasure detection that failed with a false-positive error, we either erase the qubit or completely depolarize it, each with probability $1/2$.
This captures our assumption that no coherences are created between the computational and erasure subspaces.

The rate of false positives is influenced by the measurement error rate, which is anticipated to be higher than the true erasure rate. This is because there is little motivation within the field to lower the measurement error rate below 1\%, and since the erasure rate is expected to be even lower, it is reasonable to utilize parameters where both rates are equivalent.

Since the main advantage of dual-rail qubits comes from the knowledge about the locations of erasures, the false-positive detection rate would have a noticeable effect on the QEC thresholds, as a large fraction of erasure detections would be inaccurate.
This effect would be more pronounced in the 4 EC schedule compared to the 2 EC and 1 EC schedules.

Assuming no Pauli errors, i.e., $p=0$, we find the maximal erasure rate $e$ for a range of false-positive erasure detection rates $q_\text{fn}$; see Fig.~\ref{fig:false_negative}(b).
The data can be fitted with the ansatz
\begin{equation}
	\frac{e}{e^{*}}+\frac{q_{\text{fp}}}{q_{\text{fp}}^{*}}=1\, ,
	\label{eq:false_positive_scaling}    
\end{equation}
where $q_\text{fp}^*$ is a fitting parameter.
We find that for the one-way pulse $q^*_\text{fp} \approx 0.043, 0.059, 0.074$ for the 4 EC, 2 EC and 1 EC schedules, respectively.
In contrast, for the unitary pulse $q^*_\text{fp} \approx 0.02, 0.026, 0.03$. The standard deviation of our fit was less than $1\%$ of each of these values.

\subsection{Biased erasure on dual-rail qubits}
\label{Biased_noise_XZZX}

In our simulations, we incorporate the lack of knowledge about the timing of the jump operator during the 2Q gate by fully depolarizing the partner dual-rail qubit. This approach is justified by the Hamiltonian of the 2Q gate, which is a flip-flop term of the form $X_1X_2+Y_1Y_2$ (see Ref.~\cite{kubica2023heralding} for details on dual-rail 2Q gates). Energy relaxation combined with this interaction results in both bit-flip and phase-flip errors.

A possible alternative to implementing the 2Q gate with a flip-flop interaction is to use a level repulsion between capacitively coupled dual-rail qubits, as described in Ref.~\cite{zuk2023robust}. This change to the physical operation of the gate results in a different noise profile.
If the two dual-rail qubits are not resonant, the interaction would
be along the $ZZ$ direction, evolving according to a Hamiltonian of the form
\begin{equation}
	H=\omega_{1}Z_{1}+\omega_{2}Z_{2}+g_{\text{zz}}Z_{1}Z_{2}\, ,
	\label{eq:biased_two_qubit Hamiltonian}
\end{equation}
where $Z_i$ is the Pauli $Z$ operator for the dual-rail qubit $i=1,2$.
The interaction stops when one of the qubits decays to the $\left|00\right>$ state. This decay would result only in phase errors on the dual-rail qubit that did not decay (see Appendix \ref{appx:baised_erasure}), and this noise bias can be exploited. We simulated the effect of the biased noise model on the XZZX surface code~\cite{bonilla2021xzzx,darmawan2021practical,xu2023tailored,Dua2024} and found improved thresholds of $0.0172$, $0.0175$, and $0.014$, respectively, for the 4 EC, 2 EC and 1 EC schedules. For more details on our analysis, see Appendix~\ref{appx:baised_erasure}.

\section{Discussion}
\label{sec:discussion}
In this article, we studied the performance of the surface code with erasure qubits.
We probed the threshold surfaces for different EC schedules and analyzed their subthreshold behavior.
We identified the parameter regimes in which it is optimal to use each EC schedule.
Notably, even with less precise erasure information from infrequent ECs, the surface code with erasure qubits can outperform the implementation with standard qubits.
In terms of hardware realizations of erasure qubits, we analyzed the dual-rail qubit and compared it with the standard transmon-based architecture.
While the dual-rail qubit offers an advantage in suppressing dephasing errors, its effectiveness in tackling amplitude damping errors (assuming state-of-the-art parameters) requires gates and measurements which are faster than about $200\unit{ns}$ to be below the surface code threshold. Although dual-rail transmons have not yet been demonstrated at scale, their reduced connectivity requirement---only three-qubit connectivity per transmon, compared to four in conventional surface code layouts---and strong intrinsic nonlinearity suggest that scaling them would introduce fewer correlated errors than in standard transmon-based processors.

Our study is a distinctive approach to demonstrating QEC in the lab.
Although fabricated superconducting devices are usually created for a particular QEC code, the QEC protocol can be modified based on the observed qubit lifetime, gate fidelities, and, in particular, measurement efficiencies.
For example, the achievable measurement time, following an experimental optimization of the readout procedure, should dictate the rate of ECs within the circuit.
A similar result is applicable to cavity-based dual-rail qubits, as we discuss in Appendix~\ref{appx:cavity_dual_rail}.
For erasure qubits based on neutral atoms, erasure detection is cheap enough to make the 4 EC schedule the most beneficial; see Appendix~\ref{appx:rydberg_atoms}.

While we focused on the surface code as a quantum memory, we expect other QEC protocols to similarly benefit from erasure qubits.
Future study could explore other QEC codes, such as quantum low-density parity-check codes~\cite{Breuckmann2021}, or protocols for implementing logical operations.
It would be interesting to see if the improvement from erasure qubits is universal or if certain QEC protocols particularly benefit from erasure information.

The decoding problem with erasure qubits is another direction of future explorations.
Although our work and Ref.~\cite{gu2023faulttolerant} provide ways to decode general erasure circuits by converting them to stabilizer circuits, the loss in performance due to the invoked approximations is unclear.
Decoders that directly work with erasure circuits, perhaps for specific QEC codes, may have more rigorous performance guarantees and use erasure information optimally.

Further optimization of QEC protocols with erasure qubits is possible.
One may consider independently placing ECs and reset operations throughout the circuit.
Performing multiple ECs before resetting could be useful in the presence of high false-positive erasure detection rates in order to gain more confidence in the erasure information, or if ECs can be performed concurrently with gates so that only reset operations incur noise.
Furthermore, circuits may be implemented adaptively, e.g., once an erasure is detected, we may stop extracting the syndrome of stabilizers involving that qubit to mitigate the spread of errors.
All these approaches are complementary, and incorporating them would further increase the benefits of using erasure qubits.

\acknowledgements
We thank F.~Brand\~{a}o, A.~Haim, J.~Iverson, H.~Levine and C.~Ryan for insightful discussions.
We thank A.~Grimsmo for helping correctly implement the $XZZX$ surface code. S.G. acknowledges funding from the Air Force Office of Scientific Research (FA9550-19-1-0360).

\emph{Note added.---}We would like to bring the reader’s attention to a related, independent work by Chang et al.~\cite{chang2024}, which explores the effects of physically-motivated imperfect erasure checks on the performance of the surface code and appeared in the same arXiv posting.

\appendix

\section{Example of circuit conversion for decoding}
\label{app:decodingexample}
We present an explicit example of the conversion of an erasure circuit to a stabilizer circuit, highlighting the difference between the exact conversion of Ref.~\cite{gu2023faulttolerant} and the approximate conversion of Sec.~\ref{subsec:approximatedecoding}. Consider a segment of an erasure circuit with three erasure locations and two entangling gates, as in Fig.~\ref{fig:example_erasure_circuit}(a). Assume that each erasure location has erasure rate $e$ and that the ECs have false-positive and false-negative detection rates $q$. The EC outcomes have probabilities
\begin{align}
	\Pr(\mathrm{EC}=1) &= [1 - (1 - e)^3](1 - q) + (1 - e)^3q\, ,\\
	\Pr(\mathrm{EC}=0) &= 1 - \Pr(\mathrm{EC}=1)\, .
\end{align}
If a positive detection occurs, the posterior probability of the erasure happening before gate $G_i$ is
\begin{equation}
	\bar a_i = 	\frac{(1 - (1 - e)^i)(1 - q)}{\Pr(\mathrm{EC}=1)}
\end{equation}
for $i\in \{1, 2, 3\}$ (counting $G_3$ as the second reset operation). If $\mathrm{EC} = 0$, the probability is
\begin{equation}
	\bar a_i = \frac{(1 - (1 - e)^i)q}{\Pr(\mathrm{EC}=0)}\, .
\end{equation}
Thus, in the approximate scheme, we place depolarizing channels with error probability $3\bar a_i/4$ at each location $\mc F_i$ in the circuit in Fig.~\ref{fig:example_erasure_circuit}(b). Equivalently, we place Pauli $X$, $Y$, and $Z$ error channels with strength $p_i = \frac 1 2 \left(1 - \sqrt{1 - \bar a_i}\right)$ at each location.

\begin{figure}[htpb]
	\centering
	(a) $\Qcircuit @C=.8em @R=.7em {
		& & \lstick{\ldots} & \multigate{1}{G_1} & \rstick{\ldots} \qw \\
		& \gate{\mathrm{EC^*}} & \measure{\mc E_1} & \ghost{G_1} & \measure{\mc E_2} & \multigate{1}{G_2} & \measure{\mc E_3} & \gate{\mathrm{EC^*}} & \qw\\
		& & & & \lstick{\ldots} & \ghost{G_2} & \rstick{\ldots} \qw }$\\
	\vspace*{10pt}
	(b) \qquad $\Qcircuit @C=.8em @R=.7em {
		\lstick{\ldots} & \multigate{1}{G_1} & \measure{\mc F_1} & \rstick{\ldots} \qw \\
		& \ghost{G_1} & \qw & \multigate{1}{G_2} & \qw & \measure{\mc F_3} & \qw\\
		& & \lstick{\ldots} & \ghost{G_2} & \measure{\mc F_2} & \rstick{\ldots} \qw }$
	\caption[An example of exactly or approximately converting an erasure circuit to a stabilizer circuit.]{
		(a) A segment of an erasure circuit with two entangling gates and three erasure locations.
		(b) A converted stabilizer circuit. Locations $\mc F_1, \mc F_2, \mc F_3$ are replaced with correlated depolarizing channels to obtain an equivalent stabilizer circuit or uncorrelated depolarizing channels in the approximate conversion scheme. 
	}
	\label{fig:example_erasure_circuit}
\end{figure}

In contrast, exact conversion requires capturing the correlated errors induced by erasures. If an erasure occurs at location $\mc E_i$, it would result in depolarizing at locations $\mc F_j$ for all $j\ge i$. In particular, the possibility of erasure at $\mc E_1$ requires us to insert correlated depolarizing channels at locations $\mc F_1, \mc F_2, \mc F_3$ with some probability. Decomposing this error into binary random variables requires introducing 63 error mechanisms, which correspond to the nontrivial Pauli strings of length three. For more details, see Ref.~\cite{gu2023faulttolerant}.

\section{Numerical simulation details}
\label{app:simulation_details}
In this appendix, we provide some more details on our numerical simulations and present several supplementary plots.

In our simulations, we noiselessly initialize an eigenstate of a chosen logical Pauli operator of the surface code with distance $d$.
Then, we run $3d$ rounds of the noisy syndrome extraction circuit.
Finally, we perform an ideal measurement of the logical operator. A logical error is reported if the final decoded value of the logical operator is different than when initialized. The stabilizer simulations are done using Stim~\cite{gidney2021stim}. In the decoding process, we convert the erasure circuit to a stabilizer circuit, use Stim to obtain a decoding hypergraph and further decompose the hyperedges by approximating them as independent edges, and find the minimum-weight perfect matching solution using PyMatching~\cite{higgott2023pymatching}.

In the sampling process, we first sample the erasures and erasure detection events. For each erasure sample, we then sample the Pauli errors in the circuit. The reason for doing the sampling in two stages is that each erasure sample corresponds to a different Stim circuit. As the overhead of initializing a Stim circuit is significant compared to sampling from the circuit and decoding, it is efficient to sample from the same Stim circuit multiple times. The logical error rate $p_L'$ is the total fraction of errors obtained over at least 1000 erasure samples and 200 samples of Pauli errors per corresponding circuit. We then report the normalized logical error rate per $d$ syndrome extraction rounds, calculated as $p_L = \frac 1 2(1 - (1 - 2p_L')^{1/3})\approx p_L'/3$.

\begin{figure}[ht]
	\centering
	\raisebox{50mm}{(a)}\includegraphics[width=0.45\textwidth,trim={0 0 0 0},clip]{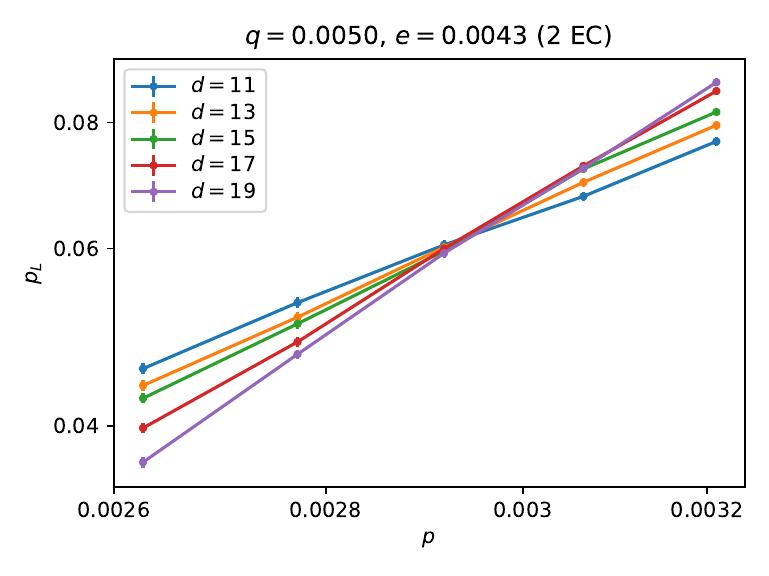}
	\raisebox{50mm}{(b)}\includegraphics[width=0.45\textwidth,trim={0 0 0 0},clip]{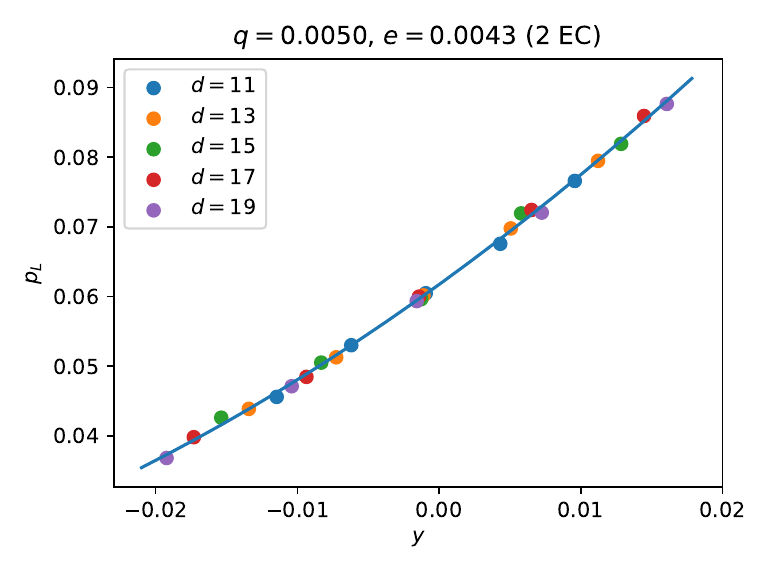}
	\caption[Sample calculation of a threshold point.]{Sample calculation of a threshold point. (a) We test values of $p$ near a suspected threshold point for fixed $e$ and $q$. (b) Rescaled data using the universal ansatz in Eq.~\eqref{eq:ansatzquadratic}, giving a threshold of $p = 0.00294$.}
	\label{fig:samplecalculation}
\end{figure}

Threshold points are calculated by sweeping a single error parameter $y\in \{e, p, q\}$ near the suspected point in $(e, p, q)$ phase space for various code distances $d$. We fit the universal scaling ansatz for critical points of phase transitions to the data,
\begin{equation}
	\label{eq:ansatzquadratic}
	p_L = ax^2 + bx + c\, ,
\end{equation}
where
\begin{equation}
	\label{eq:ansatzrescaledvariable}
	x = (y - y^*)d^{\alpha}\, .
\end{equation}
Here, $a, b, c, y^*, \alpha$ are fitting parameters, and $y^*$ is the estimated threshold. For an example calculation, see Fig.~\ref{fig:samplecalculation}.

\begin{figure*}[htpb]
	\centering
	\includegraphics[width=0.32\textwidth,trim={3mm 3mm 3mm 4mm},clip]{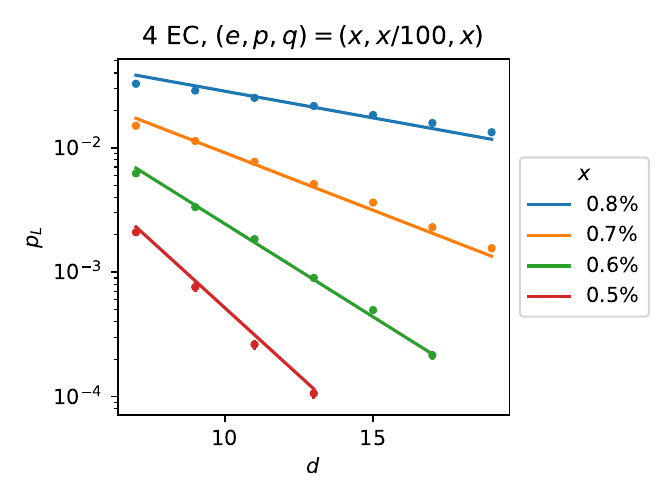}
	\hspace*{-0.32\textwidth}\raisebox{36mm}{(a)}\hspace*{0.295\textwidth}
	\includegraphics[width=0.32\textwidth,trim={3mm 3mm 3mm 4mm},clip]{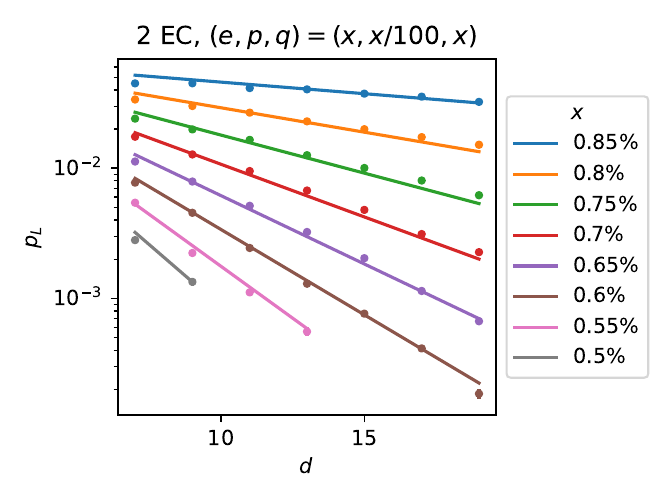}
	\hspace*{-0.32\textwidth}\raisebox{36mm}{(b)}\hspace*{0.295\textwidth}
	\includegraphics[width=0.32\textwidth,trim={3mm 3mm 3mm 4mm},clip]{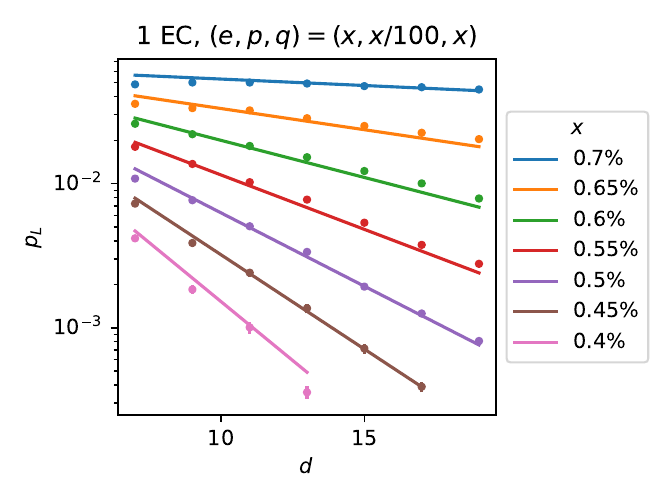}
	\hspace*{-0.32\textwidth}\raisebox{36mm}{(c)}\hspace*{0.295\textwidth}\newline
	\includegraphics[width=0.32\textwidth,trim={3mm 3mm 3mm 4mm},clip]{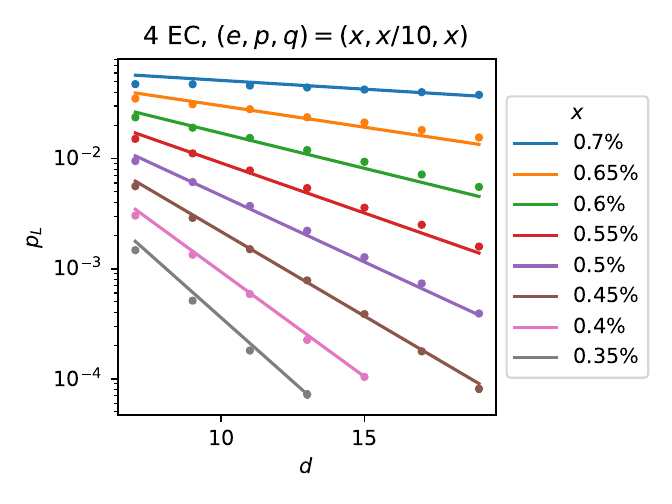}
	\hspace*{-0.32\textwidth}\raisebox{36mm}{(d)}\hspace*{0.295\textwidth}
	\includegraphics[width=0.32\textwidth,trim={3mm 3mm 3mm 4mm},clip]{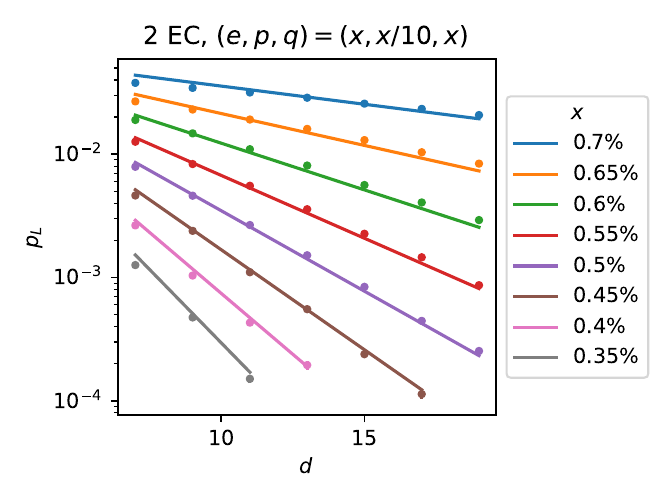}
	\hspace*{-0.32\textwidth}\raisebox{36mm}{(e)}\hspace*{0.295\textwidth}
	\includegraphics[width=0.32\textwidth,trim={3mm 3mm 3mm 4mm},clip]{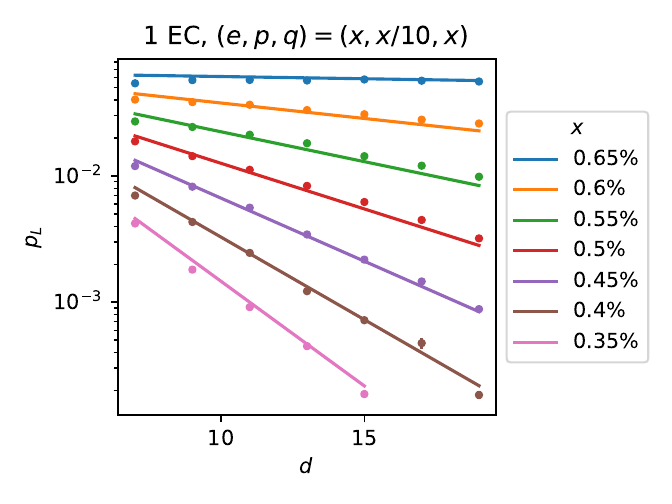}
	\hspace*{-0.32\textwidth}\raisebox{36mm}{(f)}\hspace*{0.295\textwidth}\newline
	\includegraphics[width=0.32\textwidth,trim={3mm 3mm 3mm 4mm},clip]{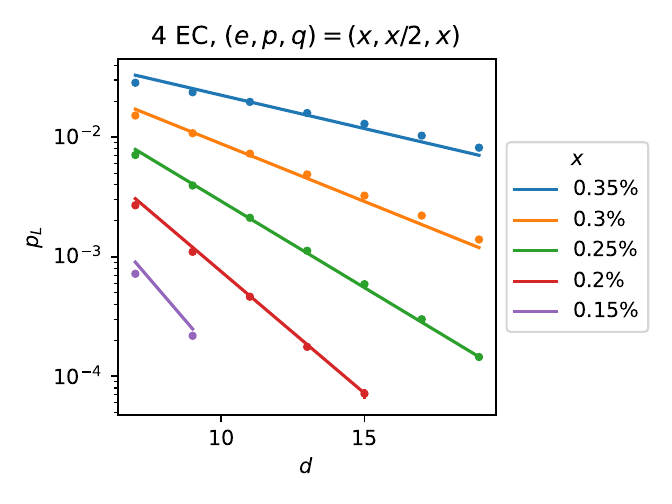}
	\hspace*{-0.32\textwidth}\raisebox{36mm}{(g)}\hspace*{0.295\textwidth}
	\includegraphics[width=0.32\textwidth,trim={3mm 3mm 3mm 4mm},clip]{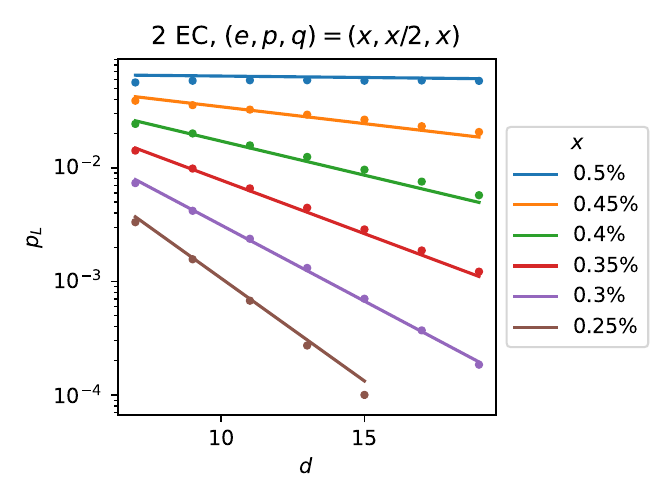}
	\hspace*{-0.32\textwidth}\raisebox{36mm}{(h)}\hspace*{0.295\textwidth}
	\includegraphics[width=0.32\textwidth,trim={3mm 3mm 3mm 4mm},clip]{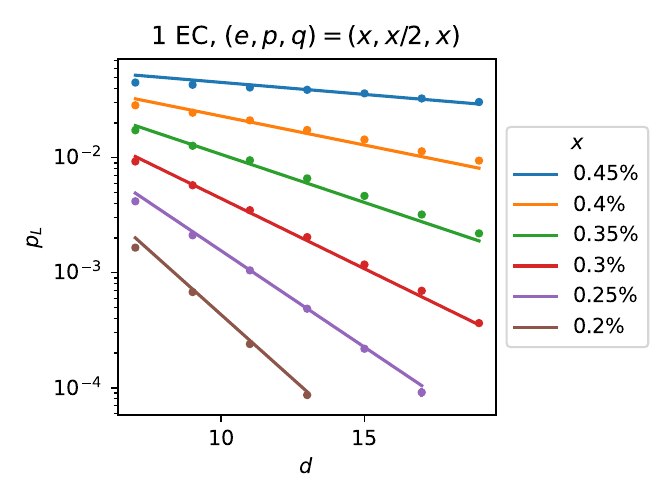}
	\hspace*{-0.32\textwidth}\raisebox{36mm}{(i)}\hspace*{0.295\textwidth}\newline
	\includegraphics[width=0.32\textwidth,trim={3mm 3mm 3mm 4mm},clip]{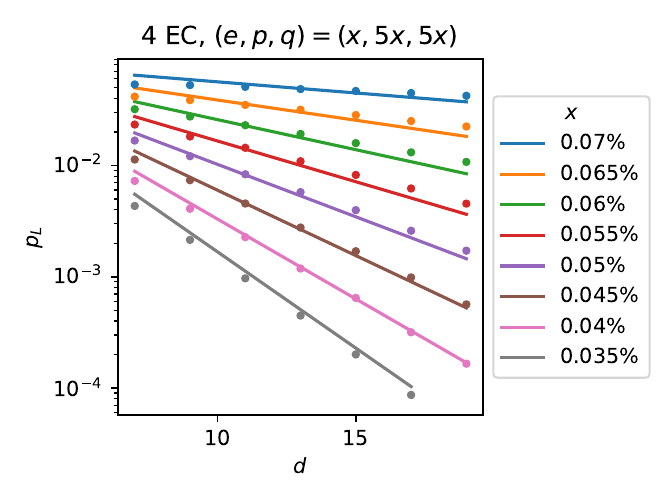}
	\hspace*{-0.32\textwidth}\raisebox{36mm}{(j)}\hspace*{0.295\textwidth}
	\includegraphics[width=0.32\textwidth,trim={3mm 3mm 3mm 4mm},clip]{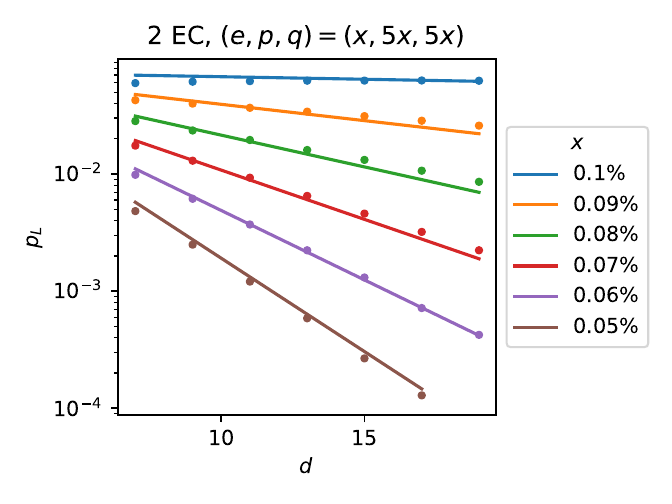}
	\hspace*{-0.32\textwidth}\raisebox{36mm}{(k)}\hspace*{0.295\textwidth}
	\includegraphics[width=0.32\textwidth,trim={3mm 3mm 3mm 4mm},clip]{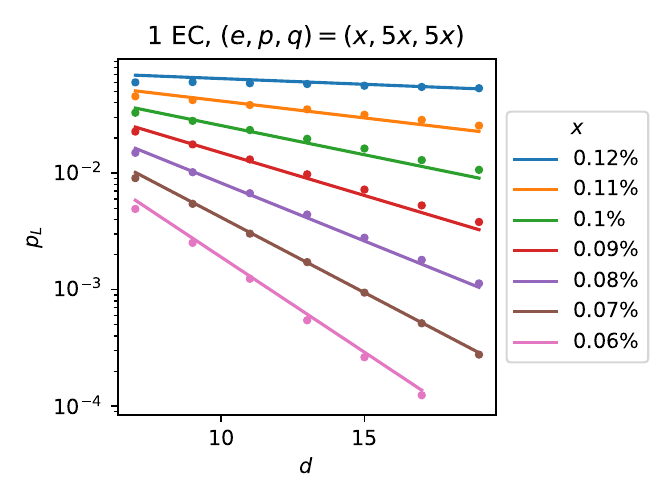}
	\hspace*{-0.32\textwidth}\raisebox{36mm}{(l)}\hspace*{0.295\textwidth}\newline
	\caption[Fitting the subthreshold logical error rate scaling ansatz of Eq.~\eqref{eq:subthreshold_ansatz} to obtain the $\alpha$ values presented in Fig.~\ref{fig:subthreshold_ansatz}(c).]{Fitting the subthreshold logical error rate scaling ansatz of Eq.~\eqref{eq:subthreshold_ansatz} to obtain the $\alpha$ values presented in Fig.~\ref{fig:subthreshold_ansatz}(c). In each plot, the different colors correspond to different values of $x$.}
	\label{fig:supplementary_subthreshold_ansatz}
\end{figure*}

Fig.~\ref{fig:supplementary_subthreshold_ansatz} shows the fit of the Eq.~\eqref{eq:subthreshold_ansatz} ansatz for the subthreshold logical error rate. The data presented are the ones used to obtain the $\alpha$ values in Fig.~\ref{fig:subthreshold_ansatz}(c).

\section{Biased erasure}
\label{appx:baised_erasure}
In this appendix, we consider in more detail our numerical results on biased erasure noise and give some additional notes on their applicability to various platforms. Using our biased noise model for the 2Q CZ gate (Eq.~\eqref{eq:baised_erasure_channel}), we estimate the effect of erasure in such systems.

Consider the 2Q gate Hamiltonian of Eq.~\eqref{eq:biased_two_qubit Hamiltonian}. Now consider one of the two dual-rail qubits decaying into its ground state. Without loss of generality, let it be the first qubit.
Then, when averaged over all possible decay times, following a reset channel on the decayed qubit, the noise channel acts on the state $\rho$ in the computational subspace of both qubits as
\begin{align}
	\mathcal{R}\left(\rho\right)&=\mathcal D_{1}\left[\frac{1}{\pi}\int_{0}^{\pi}e^{iZ_{1}Z_{2}\theta}\rho e^{-iZ_{1}Z_{2}\theta}\right] \nonumber \\
	&=\mathcal D_{1}\left[\frac{1}{2}\rho+\frac{1}{2}Z_{1}Z_{2}\rho Z_{1}Z_{2}\right] \nonumber\\
	&=\frac{1}{2}\mathcal D_{1}\left[\rho\right]+\frac{1}{2}Z_{2}\mathcal D_{1}\left[\rho\right]Z_{2}\, .
	\label{eq:baised_erasure_channel}
\end{align}
Here, $\mathcal D_1$ is a channel that fully depolarizes the first qubit and acts trivially on the second qubit, modeling the decay and reset of the decayed qubit.
Consequently, only the phase of the qubit that did not decay is lost, which effectively biases the noise.

\begin{figure}[ht]
	\centering
	\includegraphics[width=0.49\textwidth,trim={0 0 0 0},clip]{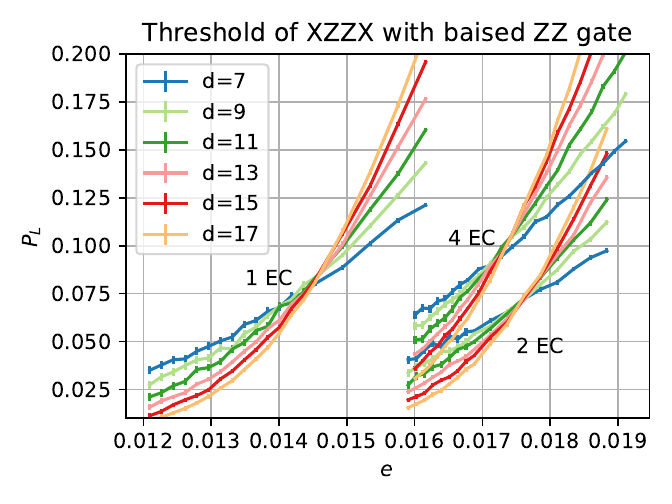}
	\caption{
		Logical error rate $p_L$ for the $XZZX$ surface code as a function of the erasure rate $e$ with for surface code of distance $d$ for $d$ rounds of QEC.
		We find threshold values $e^*\approx 0.0172, 0.0175, 0.014$ for the 4 EC, 2 EC, and 1~EC schedules, respectively.
	}
	\label{fig:supplementary_XZZX}
\end{figure}

We consider the $XZZX$ surface code~\cite{bonilla2021xzzx,darmawan2021practical,xu2023tailored,Dua2024}, which can exploit bias in the underlying noise.
In Ref.~\cite{darmawan2021practical}, a syndrome extraction circuit is designed with two $CZ$ gates and two $CX$ gates.
When implemented using $CZ$ and Hadamards, each one of the four 2Q gates during the syndrome extraction cycle can propagate a $Z$ error to a qubit when its partner is erased during the gate. Out of 8 possible $Z$ errors, two for each such gate, only two propagate as $X$ errors because of the following Hadamard gate. We use the gate schedule described in Ref.~\cite{google2023suppressing}, which suppresses any hook errors.
The improved threshold shows that the $XZZX$ surface code can exploit the bias introduced by the $ZZ$ gate; see Fig.~\ref{fig:supplementary_XZZX}.

Gates based on level repulsion are available in many quantum computing architectures~\cite{zuk2023robust, mehta2025bias} and would always give the benefit mentioned above. In systems in which this type of gate is not available or is too slow, a biased gate is still possible through a coherent excitation of a state outside the computational subspace, but such implementation requires a diagonal term of the form $Z_1 Z_2$. In neutral atoms, for example, this comes from the Rydberg blockade term $\Delta \ket{rr} \bra{rr}$.
We analyze this case in further detail in Appendix~\ref{Biased_Noise_gates}, similar to the case in Ref.~\cite{sahay2023high}.

Note the difference between this approach and Ref.~\cite{sahay2023high}. In Ref.~\cite{sahay2023high}, the erasure process similarly results in biased noise, since only a single state can decay, a property that is independent of the 2Q gate implementation.
Since only the Rydberg state can decay, any erasure detection is an effective measurement of the qubit in the computational basis. This measurement commutes with every $Z$ stabilizer, meaning that only that phase of the qubit is lost. The effect on the unerased qubit depends on the specific implementation of the gate and can result in unbiased noise and leakage (see Appendix~\ref{Biased_Noise_gates}).

\section{Biased-noise gates}
\label{Biased_Noise_gates}

In this appendix, we consider two different gates on erasure qubits that have the biased noise mechanism where the erased qubit decays to an identifiable erased state, and its partner qubit for the 2Q gate is only affected by noise in the $Z$ direction. 

As mentioned in Appendix~\ref{appx:baised_erasure}, in the case of level repulsion between two transmons or dual-rail qubits, the 2Q gate stops at an unknown time, which results in an unknown phase on the partner qubit. In this case the error on the unerased qubit is biased to a single axis, the $Z$ axis.

A second type of gate with this property is described in Ref.~\cite{levine2019parallel}, which is implemented on neutral atoms. Assume the two atoms begin in the state
\begin{equation}
	\label{eq:general_superposition}
	\ket{\psi}=a_{00}\left|00\right>+a_{01}\left|01\right>+a_{10}\left|10\right>+a_{11}\left|11\right>\, ,   
\end{equation}
where $\ket{0},\ket{1}$ are the computational states of the atom. We denote the excited Rydberg state as $\ket{r}$. 
We pulse both qubits with the following Hamiltonian:
\begin{equation}
	H=\Omega\left(\mathbb{I}\otimes\left|1\right>\left<r\right|+\left|1\right>\left<r\right|\otimes\mathbb{I}\right)+\Delta\left|rr\right>\left<rr\right|,
\end{equation}
which is written in the interaction picture with respect to the gaps:
\begin{equation}
	H_0 = \omega_0\ket{0}\bra{0}+\omega_1\ket{1}\bra{1}+\omega_r\ket{r}\bra{r}\, .
\end{equation}
In the regime of strong Rydberg blockade ($\Delta\gg\Omega$), the state evolves in the interaction picture with respect to $H_0$ according to
\begin{align}
	\left|\psi\left(t\right)\right> & =a_{00}\left|00\right>+a_{01}\left[\cos\left(\Omega t\right)\left|01\right>+i\sin\left(\Omega t\right)\left|0r\right>\right]\nonumber \\
	& +a_{10}\left[\cos\left(\Omega t\right)\left|10\right>-i\sin\left(\Omega t\right)\left|r0\right>\right]\nonumber\\
	& +a_{11}\left[\cos\left(\sqrt{2}\Omega t\right)\left|11\right>-i\sin\left(\sqrt{2}\Omega t\right)\left|W\right>\right]\, ,
\end{align}
where $\left|W\right>=\left[\left|1r\right>+\left|r1\right>\right]/\sqrt{2}$. Now assume that the first atom decayed as in Ref.~\cite{sahay2023high} from its Rydberg state $\ket{r}$ to an identifiable state $\ket{e}$. The unnormalized state  of the two atoms becomes
\begin{equation}
	\left|\psi\left(t\right)\right>=a_{10}\sin\left(\Omega t\right)\left|e0\right>+a_{11}\sin\left(\sqrt{2}\Omega t\right)\left|e1\right>/\sqrt{2}\, ,
\end{equation}
occurring at some random time $t$.  Continuing the interaction for some additional time $t_2$, the state becomes
\begin{align}
	\left|\psi\left(t+t_2\right)\right>&=-ia_{10}\sin\left(\Omega t\right)\left|e0\right>\nonumber\\&-ia_{11}\sin\left(\sqrt{2}\Omega t\right)\cos\left(\Omega t_2\right)\left|e1\right>/\sqrt{2}\nonumber\\&-a_{11}\sin\left(\sqrt{2}\Omega t\right)\sin\left(\Omega t_2\right)\left|er\right>/\sqrt{2}\, .
\end{align}
At the end of the gate, either the undecayed qubit stays in the computational manifold $\mathrm{span}\{\ket 0, \ket 1\}$ or in the Rydberg $\ket{r}$ state. In the computational manifold, notice that its transformation to the new state commutes with the $Z$ operator acting on the second atom. Hence, the undecayed atom only lost its phase, as any stabilizer with a $Z$ support on the second atom would not be affected by this transformation. The amplitude of the $\ket{er}$ state at the end of the gate describes a leakage mechanism, which does not fit into our formalism, and might present an issue for implementing biased-noise gates on neutral atoms.

\section{Error channels on dual-rail qubits and transmons}
\label{appx:error_channels}

The large benefit of detecting erasure compared with decoding Pauli noise is slightly obscured
by the way we define the Pauli and erasure rates. The erasure
probability is per-operation-per-qubit, while the Pauli rate is only per-operation, which for 2Q gates, would introduce an additional factor of 2.
To put it differently, a 2Q gate is twice as likely to fail compared
with a 1Q operation of the same length, but we give both the same
probability (which is common in the QEC literature). 

Therefore, we need to be careful when averaging the failure probability for Pauli and erasure noises. For Pauli noise, we take two qubits and consider the probability of either of them failing in an imaginary syndrome extraction cycle in which we apply a \textit{CNOT} between them four times. This cycle will include 4 CNOTs and 2$l$ measurements. The probability of either of them failing during that time is $2\times (4T_{2Q}+lT_M)/T_1$, and the average per operation, i.e. the Pauli error probability, is then
\begin{equation}
	p=\frac{8T_{2Q}+2lT_{M}}{\left(4+2l\right)T_{1}}=\frac{4T_{2Q}+lT_{M}}{\left(2+l\right)T_{1}}\, .
\end{equation}
For the erasure rate, which is a per-qubit-per-operation rate, we simply average the total cycle time over the number of operations in the cycle:
\begin{equation}
	e=\frac{4T_{2Q}+lT_{M}}{\left(4+l\right)T_{M}}\, .
\end{equation}

When comparing dual-rail qubits to transmons, the two systems are affected
by different noise channels that are not directly comparable. We now describe the channels and how we compared them in Sec.~\ref{subsec:nonerasurecomparison}.

For the dual-rail qubit, it is easiest to analyze the system using a non-Hermitian
Hamiltonian. Both transmons have some decay time $T_{1}$, and the
non-Hermitian evolution of the system is given by
\begin{equation}
	H=\frac{\Omega}{2}\left[a_{1}^{\dagger}a_{2}+\text{h.c.}\right]-i\left[\frac{1}{T_{1}^{\left(1\right)}}a_{1}^{\dagger}a_{1}+\frac{1}{T_{1}^{\left(2\right)}}a_{2}^{\dagger}a_{2}\right].
\end{equation}
We diagonalize the Hermitian part using $b_{\pm}=\left(a_{1}\pm a_{2}\right)/\sqrt{2}$,
and then move to the rotating frame with respect to $H_{0}=\frac{\Omega}{2}\left[b_{+}^{\dagger}b_{+}-b_{-}^{\dagger}b_{-}\right]$:

\begin{align}
	H_{I} &=-\frac{i}{2}\left(\frac{1}{T_{1}^{\left(1\right)}}+\frac{1}{T_{1}^{\left(2\right)}}\right)\left[b_{+}^{\dagger}b_{+}+b_{-}^{\dagger}b_{-}\right]\nonumber\\
	&-\frac{i}{2}\left(\frac{1}{T_{1}^{\left(1\right)}}-\frac{1}{T_{1}^{\left(2\right)}}\right)\left[b_{+}^{\dagger}b_{-}e^{i\Omega t}+\text{h.c.}\right].
\end{align}

In the rotating wave approximation ($\Omega\gg T_{1}^{\left(i\right)}$),
the rotating terms disappear, and both logical states decay to the ground
at an average rate
\begin{equation}
	T_{1}=\left(\frac{1}{2}\left[\frac{1}{T_{1}^{\left(1\right)}}+\frac{1}{T_{1}^{\left(2\right)}}\right]\right)^{-1}\, ,
\end{equation}
which is simply $T_{1}=T_{1}^{\left(1\right)}=T_{1}^{\left(2\right)}$ if both transmons have the same decay rate. 

\begin{figure*}[htpb]
	\centering
	\raisebox{51mm}{(a)}\includegraphics[width=0.45\textwidth,trim={0 0 0 0},clip]{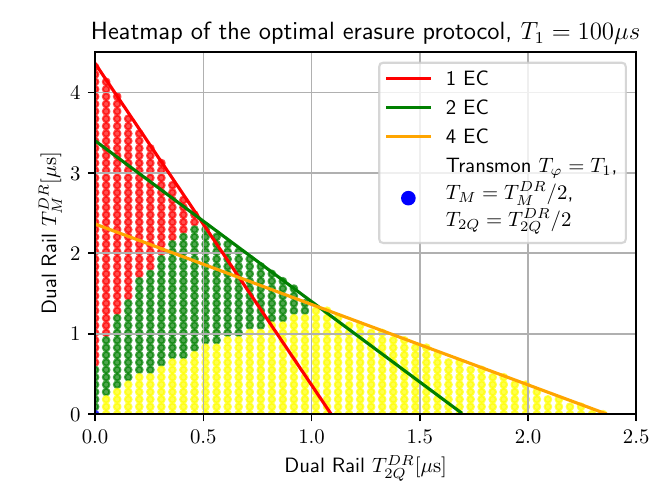}
	\raisebox{51mm}{(b)}\includegraphics[width=0.45\textwidth,trim={0 0 0 0},clip]{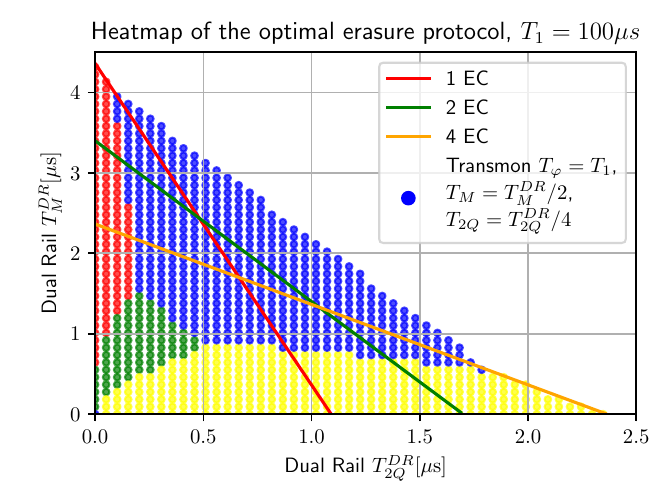}
	\raisebox{51mm}{(c)}\includegraphics[width=0.45\textwidth,trim={0 0 0 0},clip]{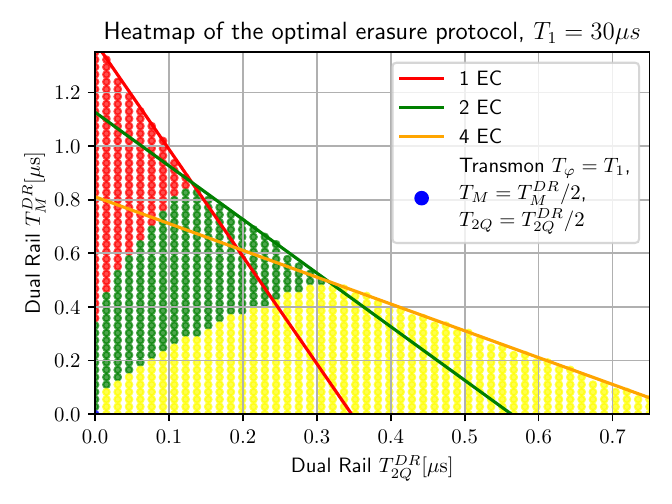}
	\raisebox{51mm}{(d)}\includegraphics[width=0.45\textwidth,trim={0 0 0 0},clip]{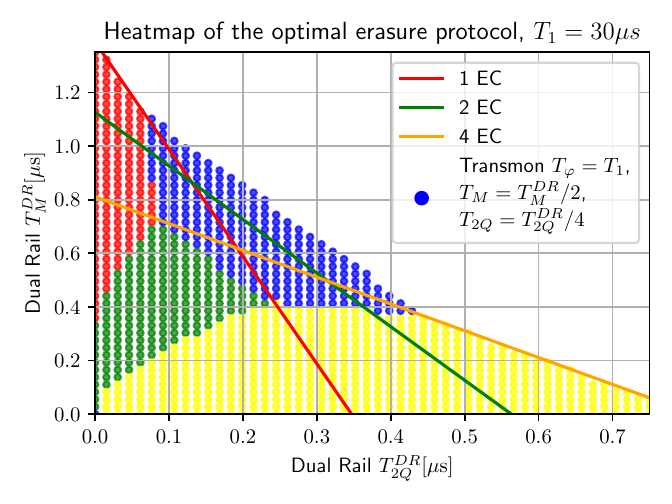}
	\caption[Heatmap of choice of optimal Protocol]{Cross-sections of the subthreshold surface for dual-rail qubits. The parameters are the same as in Fig.~\ref{fig:dual_rail_threshold}, but the colored regions represent the optimal protocol choice as a function of the dual-rail qubit's $T_{2Q}^{DR}$ and $T_M^{DR}$ times---yellow, green, and red for 4 EC, 2 EC, and 1 EC, respectively. The blue regions label points where transmons (with faster gate and measurement times) outperform the dual-rail qubit.
		The transmon $T_{2Q}$ and $T_M$ values scale linearly with these parameters, as labeled in the legends.
		The lines represent error-correcting thresholds for each EC protocol. In panels (a) and (b) we set $T_1=100\mu s$, while in panels (c) and (d) $T_1$ is reduced to $30\mu s$, resulting in a greater advantage for the dual-rail qubit.}
	\label{fig:dual_rail_threshold_heatmap}
\end{figure*}

To analyze the decay of the transmon, we look at its energy relaxation
Linbdladian
\begin{equation}
	\mathcal{L}\left(\rho\right)=\frac{1}{T_{1}}\left(a^{\dagger}\rho a-\frac{1}{2}\left[a^{\dagger}a\rho+\rho a^{\dagger}a\right]\right)\, .
\end{equation}
Applied for time $t$, we have
\begin{equation}
	U_{T_{1}}\left(\rho\right)=\left(\begin{matrix}\rho_{11}e^{-t/T_{1}} & \rho_{12}e^{-t/2T_{1}}\\
		\rho_{12}^{*}e^{-t/2T_{1}} & \rho_{22}+\rho_{11}\left(1-e^{-t/T_{1}}\right)
	\end{matrix}\right)\, .
\end{equation}
This channel cannot be directly approximated by a Pauli channel, but we can reduce it to a Pauli channel using the Pauli twirling approximation~\cite{geller2013efficient}. In this form, the twirled channel is
\begin{align}U_{T_{1}}^{P}\left(\rho\right) & =\left(e^{-t/2T_{1}}+\frac{1}{2}e^{-t/T_{1}}-\frac{1}{2}\right)\rho\nonumber\\
	& +\frac{1}{4}\left(1-e^{-t/T_{1}}\right)\left[\frac{1}{2}\sigma_{x}\rho\sigma_{x}+\frac{1}{2}\sigma_{y}\rho\sigma_{y}\right]\nonumber\\
	& +\left(\frac{1}{2}-\frac{1-e^{-t/T_{1}}}{4}-\frac{e^{-t/2T_{1}}}{2}\right)\sigma_{z}\rho\sigma_{z}\, .
\end{align}
For short times $t\ll T_{1}$, we get
\begin{equation}
	U_{T_{1}}^{P}\left(\rho\right)\approx\left(1-t/2T_{1}\right)\rho+\frac{t}{2T_{1}}\left[\frac{1}{2}\sigma_{x}\rho\sigma_{x}+\frac{1}{2}\sigma_{y}\rho\sigma_{y}\right]\, .
\end{equation}
This is a slightly biased channel along the $X,Y$ direction. In order
to approximate this channel as a depolarizing channel of the form
\begin{equation}
	U_{p}\left(\rho\right)=\left(1-p\right)\rho+\frac{p}{3}\sum_{a\in\left\{ x,y,z\right\} }\left(\sigma_{a}\rho\sigma_{a}\right)\, ,
\end{equation}
we assign $p=t/2T_{1}$ as the probability of any Pauli error.

We model dephasing using the following Markovian Lindblad operator:
\begin{equation}
	\mathcal{L}\left(\rho\right)=\frac{1}{2T_{\varphi}}\left[\sigma_{z}\rho\sigma_{z}-\rho\right]\, .
\end{equation}
The channel after time $t$ is
\begin{equation}
	U_{T_{\varphi},t}\left(\rho\right)=\left(\frac{1}{2}+\frac{1}{2}e^{-t/T_{\varphi}}\right)\rho+\left(\frac{1}{2}-\frac{1}{2}e^{-t/T_{\varphi}}\right)\sigma_{z}\rho\sigma_{z}\, .
\end{equation}
For short times $t\ll T_{\varphi}$,
\begin{equation}
	U_{T_{\varphi},t}\left(\rho\right)\approx\left(1-\frac{t}{2T_{\varphi}}\right)\rho+\frac{t}{2T_{\varphi}}\sigma_{z}\rho\sigma_{z}\, .
\end{equation}

\section{Choice of optimal protocol}
\label{choice_of_optimal_protocol}

Given a set of values for a particular system $T_1$, $T_\varphi$, $T_M$, and $T_{2Q}$, it is natural to ask what the optimal choice of EC protocol is to achieve the lowest logical error rate. We estimate this value using Eq.~\eqref{eq:subthreshold_ansatz} by linearly interpolating the values of $\alpha$ from Fig~\ref{fig:subthreshold_ansatz}(c) as a function of $\log t_p$.  Given this interpolation, we can estimate how the error would scale as a function of $d$. A protocol is considered \textit{optimal} if it gives the minimal value of $(x/x^*)^\alpha$. 

The results are summarized in Fig \ref{fig:dual_rail_threshold_heatmap}. We observe that by drawing a straight line from the origin to any point within the correctable region, the optimal protocol for that point is typically the one with the threshold farthest along that line.

When the gate and measurement times for dual-rail and transmon qubits are the same, the entire parameter space is dominated by the erasure detection protocol. Even when transmon gates \emph{and} measurements are $2\times$ faster, the dual-rail qubit still outperforms it. Only if the transmon gates are $4\times$ faster do we see a region where the transmons outperform the dual-rail qubit. We plot two values for the single transmon $T_1 \in \{100\mu s, 30\mu s\}$. Since we set the transmon's dephasing time to be $T_1$, the dual-rail qubit improves the scaling of the logical error rate for lower $T_1$. 

\section{One way pulse derivation}
\label{one_way_pulse_derivation}
The double excitation subspace of the dual-rail qubit is spanned by a dark
state $\left|D\right>=\sqrt{1/2}\left[\left|02\right>-\left|20\right>\right]$
and two additional bright states. The internal dual-rail qubit coupling
couples the state $\left|B\right>=\sqrt{1/2}\left[\left|02\right>+\left|20\right>\right]$
to the state $\left|S\right>=\left|11\right>$ with the following
interaction:
\begin{equation}
	H=\left(\begin{matrix}2\omega_{DR}+\eta & 2g_{12}\\
		2g_{12} & 2\omega_{DR}
	\end{matrix}\right)\, .
\end{equation}
The eigenstates of this Hamiltonian have energies $E_{\pm}=2\omega_{DR}+\frac{\eta}{2}\pm\sqrt{\frac{\eta^{2}}{4}+4g_{12}^{2}}$
which we label $\left|B_{\pm}\right>$. The transition to the lowest
energy state $\left|B_{+}\right>$ is the most resonant during the
reset pulse, giving the shift
\begin{equation}
	\Delta=E_{\left|B_{+}\right>\otimes\left|1\right>}-E_{\left|\bar{1}\right>\otimes\left|0\right>}-\omega_{d}=\frac{\eta}{2}\pm\sqrt{\frac{\eta^{2}}{4}+4g_{12}^{2}}\, .
\end{equation}

\section{Other platforms}
\subsection{Cavity dual-rail qubit}
\label{appx:cavity_dual_rail}
When considering a cavity dual-rail qubit, a general assumption is that the limiting factor $T_1^C$ is the cavity amplitude damping, which is dictated by the intrinsic decay and the Purcell effect, while coherence is influenced by the transmon and is fairly small. The erasure rate during a 2Q gate is $e = \frac{T_{2Q}}{T_1^C}=\frac{\Delta^2}{g^2}\frac{1}{\alpha T_1^C}$ . The limit on erasure, however, will be dominated by the erasure rate of the transmon $e^T = \frac{T_{2Q}}{T_1^T}=\frac{\Delta^2}{g^2}\frac{1}{\alpha T_1^T}>e^C$ and the dephasing erasure rate on the transmon $e^T_\phi = \frac{T_{2Q}}{T_\phi^T}=\frac{\Delta^2}{g^2}\frac{1}{\alpha T_\phi^T}>e^T$. As dephasing does not propagate under dispersive interaction, its main effect would be to reduce the measurement fidelity, and thus, the main erasure bottleneck would be dominated by $e_T.$  As it can be assumed that we are in the regime in which $T_1 \ll T_\phi$ and $T_\phi$ is smaller than the transmon-induced $T_\phi$, the bias is $\frac{T_\phi}{T_1^T}$, which should be large enough to allow the benefits of the doubling of the erasure distance, which is between 10 to 100; see Fig.~\ref{fig:subthreshold_ansatz}(c).

As reported in Ref.~\cite{koottandavida2024erasure}, the cavity amplitude damping time is $T_1 \approx 250 \mu\text{s}$, and the intrinsic cavity dephasing is very long. The gate time is around $2 \mu\text{s}$, resulting in an erasure cavity error of $0.8\%$. The erasure ancilla error is $1.3 \%$ as its $T_1$ is $147 \mu\text{s}$.  The measurement time is around $12 \mu\text{s}$, which results in erasure rate of $4.8\%$ during the measurement. 

\begin{figure}[htpb]
	\centering
	\includegraphics[width=0.45\textwidth,trim={0 0 0 0},clip]{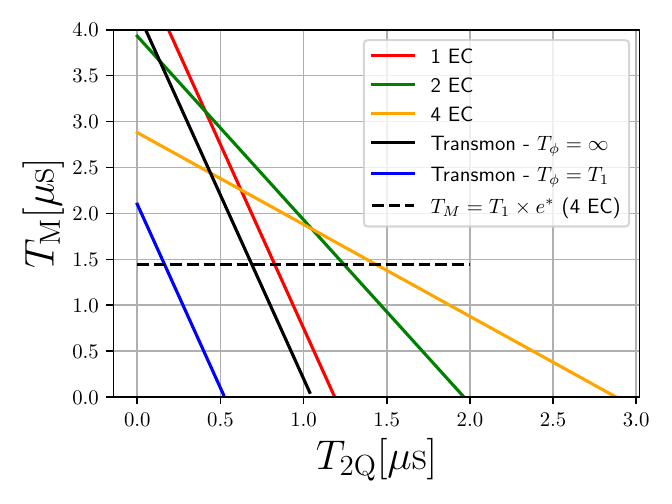}
	\caption[Cavity]{The $(T_1, T_\phi)$ threshold curves. The dephasing time is chosen to be long enough to be negligible. }
	\label{fig:Cavity}
\end{figure}

The subthreshold region is shown in Fig.~\ref{fig:Cavity} as a function of the measurement time and the 2Q gate time. As the results are comparable to the transmon dual-rail qubit (Fig.~\ref{fig:dual_rail_threshold}), the cavity dual-rail qubit operations are not fast enough currently to show subthreshold behavior. 

\subsubsection{Comparison to the transmon dual-rail qubit}
One major advantage of the cavity dual-rail qubit over the transmon variant is the improved measurement fidelity, attributed to the longer cavity $T_1$. Additionally, switching from a single cavity to a dual-rail qubit enables erasure detection and 1Q operations, while in the case of the transmon, it also significantly extends dephasing time.

Unlike the transmon dual-rail qubit, there is no single-rail analogue for the cavity, as the transition $\ket 0 \leftrightarrow \ket 1$ is extremely challenging for a single cavity.

\begin{figure}[ht!]
	\centering
	\raisebox{47mm}{(a)}\includegraphics[width=0.45\textwidth,trim={0 0 0 0},clip]{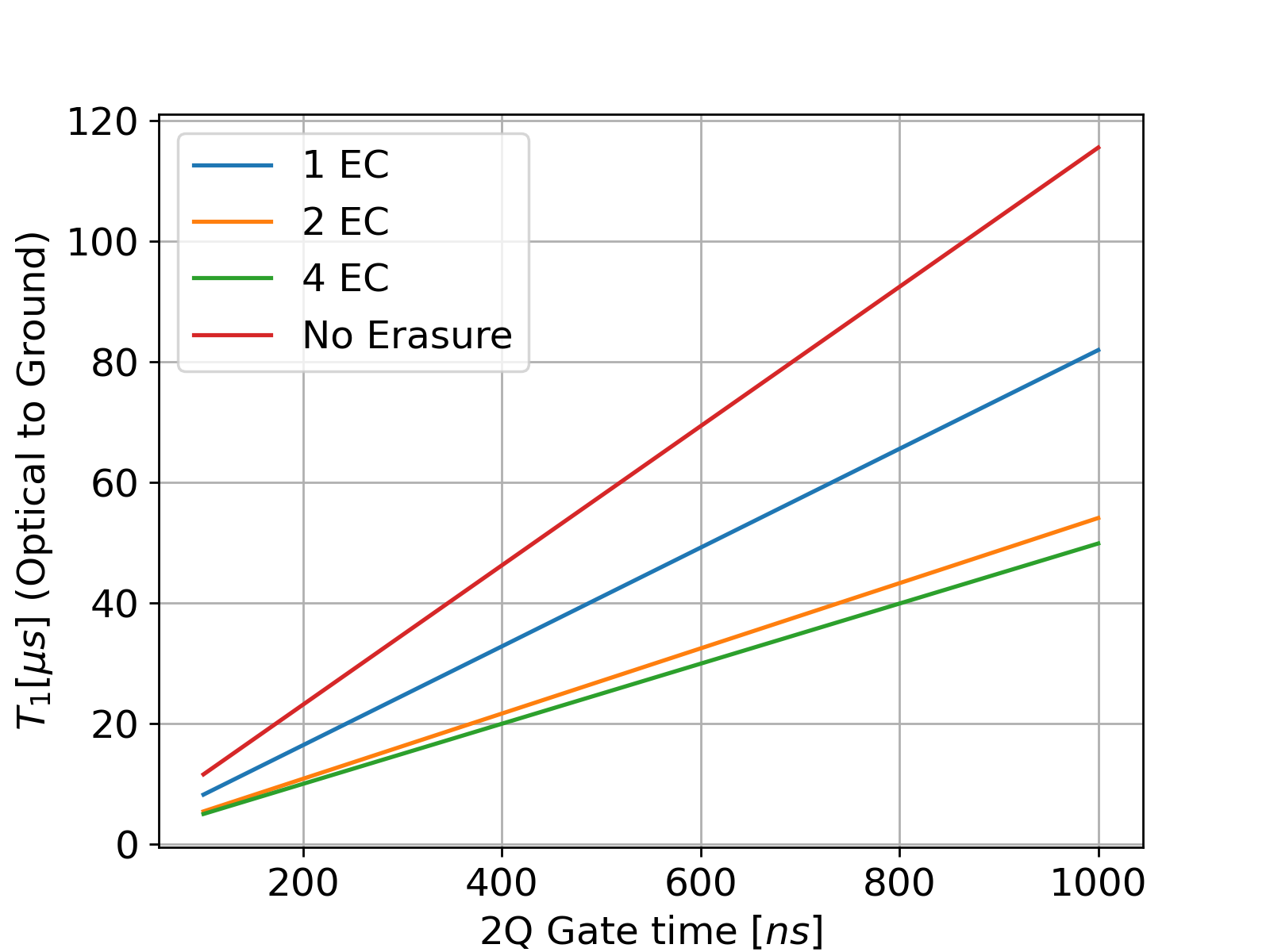}
	\raisebox{47mm}{(b)}\includegraphics[width=0.45\textwidth,trim={0 0 0 0},clip]{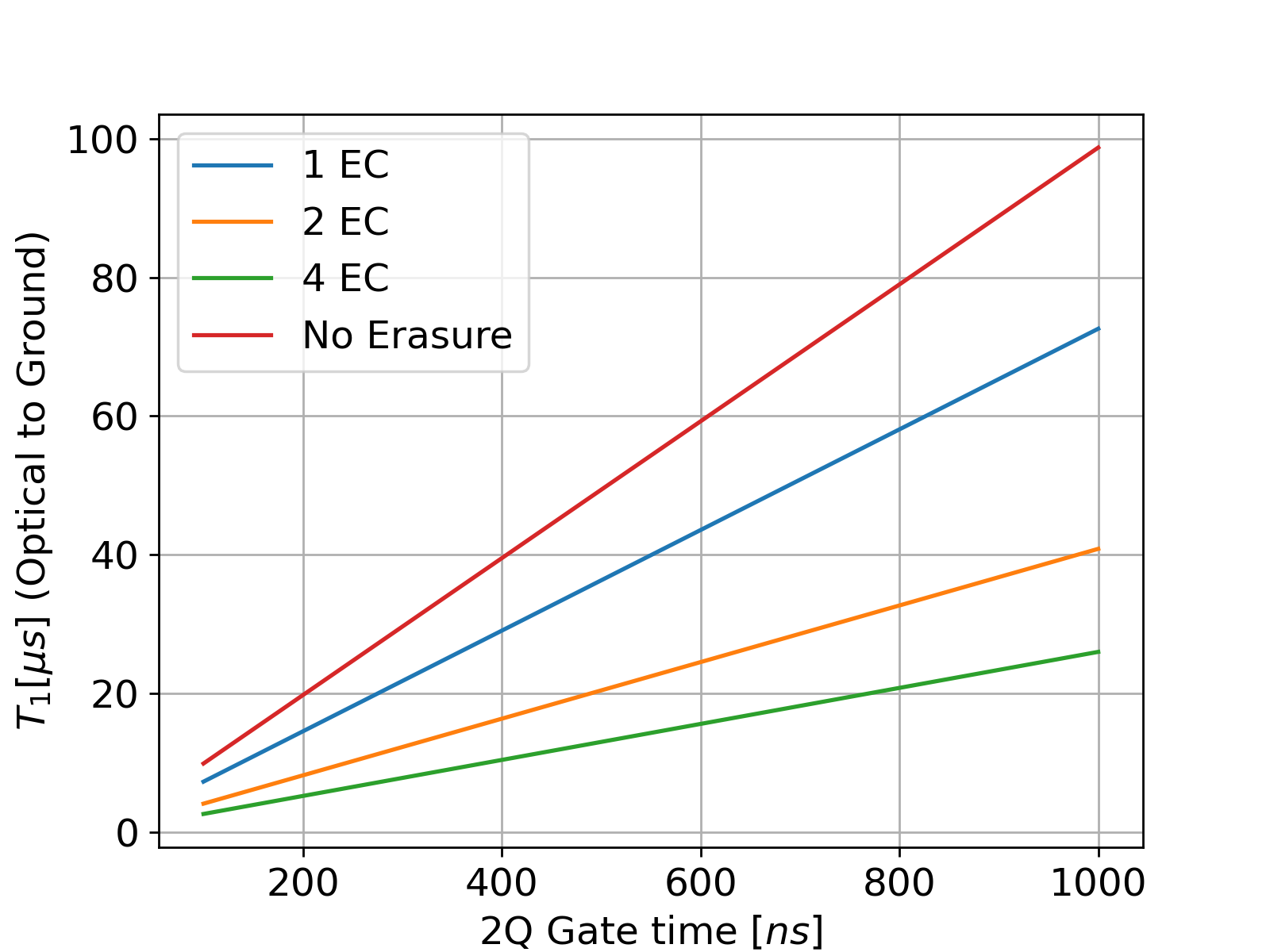}
	\caption[Rydberg]{
		The minimal required $T_1$ of the metastable manifold as a function of the 2Q gate time to reach subthreshold behavior. We assume that the bias between decay from the optical states to the ground state (and other detectable Rydberg states) is 50 times stronger than the decay back to the metastable subspace, introducing a $50\times$ bias between Pauli and erasure noise during the 2Q gate. We used 3ms as the decay time from the metastable state to the ground state. In the “No-Erasure” simulation, we considered both decay to the metastable states and to the ground states as non-detectable Pauli noise. (a) The measurement is assumed to take $20\mu\text{s}$, as in Ref.~\cite{ma2023high}. (b) The measurement is assumed to take $2\mu\text{s}$. The measurement time does make a difference in this regime, especially when comparing 2 and 4 ECs.
	}
	\label{fig:Rydberg_QEC}
\end{figure}

\subsection{Rydberg atoms}
\label{appx:rydberg_atoms}

In this section, we analyze the threshold as a function of the number of ECs in the setting of Rydberg atoms.  According to Ref.~\cite{wu2022erasure}, an erasure conversion fraction of 0.98 could theoretically be attained, and we follow this bias in our analysis. This results in a maximum bias of 50, which could potentially be sufficient to achieve a significant distance; see Fig.~\ref{fig:subthreshold_ansatz}. Measurement-induced dephasing could in principle be as low as $10^{-5}$~\cite{ma2023high}, though currently this was only achieved for the decay of the metastable state and not the Rydberg one. It is reasonable to assume that the coherence time of the qubit could be prolonged enough to not pose a limitation, and thus, the bias would only be limited by the selection rules, which justifies the use of the $50\times$ bias used in Fig.~\ref{fig:Rydberg_QEC}. The 2Q gate would be limited by power and control limitations, i.e., the faster the gate is, the higher is the Pauli error. In Fig.~\ref{fig:Rydberg_QEC}, we assume a negligible Pauli error, which is only dictated by the erasure conversion efficiency (50) and an erasure rate which is limited by the Rydberg state lifetime (assuming $100 \mu\text{s}$).

Fig.~\ref{fig:Rydberg_QEC} indicates a consistent gain due to erasure which results from the low cost of measurement in this setting. However, measurement time does make a difference as can be seen by comparing Fig.~\ref{fig:Rydberg_QEC}(a) and Fig.~\ref{fig:Rydberg_QEC}(b).

\section{Erasure biased channel}
In the following we derive the exact expression of the erasure biased channel of Eq.~\eqref{eq:baised_erasure_channel}.
\begin{align}
	\mathcal{R}\left(\rho\right)&=\mathcal D_{1}\left[\frac{1}{\pi}\int_{0}^{\pi} 
	\frac{1}{T_1-e^{-\frac{T_g}{T_1}}}e^{-\frac{t}{T_1}} e^{iZ_{1}Z_{2}\theta}\rho e^{-iZ_{1}Z_{2}\theta}\right]\nonumber\\
	&=\mathcal D_{1} [\frac{1}{\pi}\int_{0}^{\pi} 
	\frac{1}{T_1-e^{-\frac{T_g}{T_1}}}e^{-\frac{t}{T_1}} \times\nonumber\\
	&\left(\mathbb{I} \cos\theta  + i Z_1 Z_2 \sin\theta \right)
	\rho \left(\mathbb{I} \cos\theta  + i Z_1 Z_2 \sin\theta \right) ]
\end{align}
Calculating this integral in the limit $\frac{T_g}{T_1} \ll 1$ gives

\begin{align}
	\frac{1}{2}\left(\mathcal D_{1}[\rho] (1 +  \epsilon)+ (1 -  \epsilon) Z_{2}\mathcal D_{1}[\rho] Z_{2}\right)\, ,
\end{align}
where $\epsilon = \frac{1}{8 \pi ^2
} \left( \frac{Tg}{
	T_1}\right)^2.$ 

\bibliography{bib_combined}

\end{document}